\documentclass[12pt]{article}
\usepackage{amsfonts}
\usepackage{amssymb}
\usepackage{amsmath}
\usepackage{amsthm}
\usepackage{color}
\usepackage{graphics,graphicx,subfigure}
\usepackage{epsfig,epstopdf,amssymb,latexsym,verbatim}
\usepackage{graphicx}
\usepackage{multirow}
\usepackage{multicol}
\usepackage{float}
\usepackage{hyperref}
\usepackage{natbib}
\usepackage{setspace}
\usepackage{paralist}
\usepackage{algorithm}
\usepackage[noend]{algorithmic}
\usepackage{url} % not crucial - just used below for the URL 

\usepackage{tikz}
\usetikzlibrary{shapes}
\usetikzlibrary{positioning,arrows}
\usetikzlibrary{shapes,shapes.geometric,arrows,fit,calc,positioning,automata}
\usetikzlibrary{shapes,arrows,shadows}

\newcommand{\blind}{0}
\floatstyle{ruled}
\newfloat{algorithm}{tbp}{loa}
\floatname{algorithm}{Algorithm}

\newtheorem{theorem}{{\bf Theorem}}
\newtheorem{assumption}{Assumption}
\newtheorem{structure}{\bf Structure}

\def\0{{\bf 0}}
\def\A{{\bf A}}
\def\a{{\bf a}}
\def\B{{\bf B}}

\def\C{{\bf C}}

\def\D{{\bf D}}
\def\e{{\bf e}}
\def\E{{\bf E}}

\def\I{{\bf I}}

\def\P{{\bf P}}
\def\Q{{\bf Q}}
\def\U{{\bf U}}
\def\V{{\bf V}}
\def\R{{\bf R}}

\def\u{{\bf u}}

\def\X{{\bf X}}
\def\J{{\bf J}}
\def\x{{\bf x}}

\def\Y{{\bf Y}}
\def\y{{\bf y}}
\def\Z{{\bf Z}}

\def\beps{{\boldsymbol \epsilon}}

\def\bSig{{\bf \Sigma}}

\def\bG{{\boldsymbol \Gamma}}

\def\bPsi{{\boldsymbol \Psi}}
\def\bPhi{{\boldsymbol \Phi}}

\def\trans{^{\rm T}}
\def\ntrans{^{-\rm T}}
\def\wh{\widehat}

\begin{document}

\def\spacingset#1{\renewcommand{\baselinestretch}%
{#1}\small\normalsize} \spacingset{1}

\if0\blind
{
  \title{Multivariate Functional Regression via Nested Reduced-Rank Regularization}
\author{Xiaokang Liu$^1$, Shujie Ma$^2$, Kun Chen$^1$\thanks{Corresponding author; kun.chen@uconn.edu}\\
$^1$\textit{Department of Statistics, University of Connecticut}\\%[2pt]
$^2$\textit{Department of Statistics, University of California, Riverside}
}

  \maketitle
} \fi

\if1\blind
{
  \bigskip
  \bigskip
  \bigskip
  \title{Multivariate Functional Regression via Nested Reduced-Rank Regularization}

%   \begin{center}
%     {\LARGE\bf Multivariate Functional Regression via Nested Reduced-Rank Regularization}
% \end{center}
  %\medskip
  \maketitle
} \fi

\bigskip
\begin{abstract}
  % fewer than 200 words
  We propose a nested reduced-rank regression (NRRR) approach in fitting regression model with multivariate functional responses and predictors, to achieve tailored dimension reduction and facilitate interpretation/visualization of the resulting functional model. Our approach is based on a two-level low-rank structure imposed on the functional regression surfaces. A global low-rank structure identifies a small set of latent principal functional responses and predictors that drives the underlying regression association. A local low-rank structure then controls the complexity and smoothness of the association between the principal functional responses and predictors. Through a basis expansion approach, the functional problem boils down to an interesting integrated matrix approximation task, where the blocks or submatrices of an integrated low-rank matrix share some common row space and/or column space. An iterative algorithm with convergence guarantee is developed. We establish the consistency of NRRR and also show through non-asymptotic analysis that it can achieve at least a comparable error rate to that of the reduced-rank regression. Simulation studies demonstrate the effectiveness of NRRR. We apply NRRR in an electricity demand problem, to relate the trajectories of the daily electricity consumption with those of the daily temperatures. %Supplementary materials are available online.

\end{abstract}

\noindent%
%3 to 6 keywords, that do not appear in the title
\vfill

\newpage
\spacingset{1.5} % DON'T change the spacing!

\section{Introduction}

Multivariate functional data, which are generated when multiple variables are observed over certain continuum, become increasingly prevalent nowadays, partly due to the rapid advances in record keeping, inspection, and monitoring technologies in various fields. An object might be captured by cameras/scanners at a sequence of different angles/positions. The progression of a disease, as measured by various physiological indicators, may be monitored frequently over time. With the richness of such data, it is often of interest to study the association between some multivariate functional responses and predictors. For example, with half-hourly observations on temperature and electricity consumption of the city Adelaide, the interest is to explore the predictive association between the daily electricity profiles and the daily temperature profiles, for each day in a week simultaneously. Such a predictive model can then be used to infer future weekly power demand curves based on temperature forecasts, to facilitate power supply and peak load management.  %The analysis technique to investigate the functional dependence relationship between two sets of functional variables and the incorporation of potential correlation among functional responses make possible a more accurate, comprehensive and time-specific forecasting. 

%with the selected rank being the number of principal components or the pair of canonical variates, respectively

The aforementioned problem can be cast under the framework of functional regression, which has attracted considerable research efforts in the past. \citet{cardot1999functional,cardot2003spline} considered the case of regressing a scalar response variable on a functional predictor, and \citet{james2002} generalized it to the generalized linear regression setting. \citet{faraway1997} and \citet{chiou2003} derived methods for modeling univariate functional response with scalar predictors. For the case of relating a functional response and a functional predictor, \citet{yao2005functional} considered a model based on functional principle component analysis (FPCA). \citet{he2010functional} studied a model which connects functional regression to functional canonical correlation analysis (FCCA). \citet{Ebaid2008} imposed a low-rank structure on the coefficient surface and showed that low-rank regularization is closely connected to FPCA and FCCA. Extensions to the cases of multiple scalar or functional responses/predictors have been studied by various authors, e.g., \cite{matsui2008}, %\citet{matsui2009}, 
\cite{ZHU2017}, and \citet{miro2017}. Recently \citet{he2018dimensionality} proposed a multivariate varying-coefficient model to study changing effects of predictors on responses, in which FPCA is used to reduce the number of unknown coefficient functions. As for the most general situation where both the response and the predictor are multivariate and functional, \citet{Ebaid2008} considered imposing a low-rank structure on the coefficient surface with basis expansion. \citet{CHIOU2016301} incorporated into their model the possible relationship between components of responses and predictors, respectively, by conducting multivariate FPCA to two sets of variables as the first step. For a comprehensive account of functional regression, see, e.g., \citet{Morris2015} and \citet{WangChiou2016}.

We consider the general scenario where both the response and the predictor are multivariate and functional. To formulate, let $\y(t)=[y_1(t),\ldots,y_d(t)]\trans$ be a $d$-dimensional vector of zero-mean functional response with $t\in \mathcal{T}$ and $\x(s)=[x_1(s),\ldots,x_p(s)]\trans$ a $p$-dimensional vector of zero-mean functional predictor with $s\in \mathcal{S}$. We consider the multivariate functional linear regression model
\begin{align}
\y(t) = \int_{\mathcal{S}} \C_0(s,t)\x(s) ds + \beps(t),\qquad t\in \mathcal{T},\label{eq:model}
\end{align}
where $\C_0(s,t)=[c_{k,l}(s,t)]_{d\times p}$ consists of unknown bivariate functions $c_{k,l}(s,t)$ assumed to be square integrable, i.e., $\int_{\mathcal{T}} \int_{\mathcal{S}} c^2_{k,l}(s,t) dsdt<\infty$, $k=1,\ldots,d$, $l=1,\ldots,p$, and $\beps(t)$ is a $d$-dimensional zero-mean random error function. This formulation is a natural extension of the classical functional linear model (FLM) developed for univariate time-dependent responses.  %On the other hand, 
The key is on how to jointly estimating the many functional surfaces in Model \eqref{eq:model} by utilizing the potential associations among the functional variables.

%In particular, we propose and study an interesting \textit{nested reduced-rank matrix structure} that arises naturally from the functional data setup. 

In this paper, our focus is on exploring the potentials of the \textit{reduced-rank methodology} for fitting Model \eqref{eq:model} with finite samples. In classical multivariate regression, low-rank models have been commonly applied to invoke information sharing among the correlated responses and predictors, in order to boost predictive performance and enhance model interpretation \citep{izenman1975,reinsel1998,bunea2011,Chen2013ann}. It appears straightforward to utilize this idea for functional regression, once a pragmatic basis expansion/truncation procedure \citep{ramsay2005} is applied to transform the functional problem to finite dimensions. Imposing a low-rank structure on the resulting coefficient matrix is then a natural and somewhat generic choice for controlling model complexity \citep{Ebaid2008}. However, we argue that such a naive reduced-rank implementation does not take full advantage of the multivariate and functional nature of the problem, and hence can be inadequate in practice.

%In particular, in the presence of high-dimensional and highly-correlated functional variables, their underlying association may be driven through a much smaller set of ``principal'' functional responses and predictors, which enables the potential of a further dimension reduction.

%resulted from such kind of structures. For example, in the Adelaide electricity consumption data, we may expect that seven component curves of the electricity consumption corresponding to each day in a week can be summarized pretty well by only two latent functional factors, one for weekday and another for weekend. This may bring a dramastic dimension reduction and thus a better estimation.  

%After the expansion of functions and the regression surface, the reduced rank structure which is imposed on the matrix of bivariate regression functions can be viewed as the restriction on the corresponding matrix of coefficients of basis expansions. 

%In particular, we propose and study an interesting \textit{nested reduced-rank matrix structure} that arises naturally from the functional data setup. 

We innovate a \textit{nested reduced-rank matrix representation}, to enable multi-scale learning in Model \eqref{eq:model}. At the global level, our method identifies latent principal functional factors that drive the functional association between the responses and the predictors. As such, dimension reduction is achieved when the number of latent responses is less than $d$ and/or the number of latent predictors is less than $p$. This reduction can be quite effective in the presence of high-dimensional and highly-correlated functional variables. At the local level, the smaller-dimensional latent regression surface is assumed to be smooth and correspondingly its coefficient matrix derived through basis expansion is assumed to be of low rank, enabling another chance of dimension reduction. With these structures, the problem then boils down to a high-dimensional matrix decomposition and approximation task, where the nested reduced-rank structure implies that the blocks or submatrices of an integrated high-dimensional low-rank matrix share some common row space and/or column space. The applicability of the nested reduced-rank structure goes well beyond the functional setup; for instance, it also arises in vector autoregressive modeling of time series. 

The paper is organized as follows. Section \ref{NRRR} introduces the nested reduced-rank formulation under Model \eqref{eq:model}, derives the model estimation procedure, and showcases the applicability of such nested reduced-rank matrix recovery in time series modeling and image compression. Computational algorithms and rank selection methods are proposed in Section \ref{computation}. In Section \ref{theory}, we show the consistency of the proposed estimator and derive a non-asymptotic error bound. Simulation studies and the application on electricity demand are presented in Sections \ref{simu} and \ref{AppAde}, respectively. In Section \ref{diss}, we conclude with some remarks. %The proof of the theorem is included in appendix and some of the simulation and application results are listed in supplementary materials.

\section{Nested Reduced-Rank Regression}\label{NRRR}

\subsection{Model Formulation}

%A natural way to avoid overfitting using model (\ref{eq:model}) is to reduce its dimensionality by imposing certain meaningful lower-dimensional structures or constraints on the regression surface $\C(s,t)$.

%In the context of multivariate regression, it is well known that assuming a reduced-rank structure on the coefficient matrix can effectively achieve dimension reduction and account for the dependence structure among the responses.

%Inspired by various matrix decomposition and dimension reduction approaches, in particular the rank reduction methodology \citep{izenman1975,reinsel1998,bunea2011,Chen2013ann},

We propose a nested reduced-rank structure under Model \eqref{eq:model}, to appreciate both the multivariate and the functional natures of the problem.

%In Assumptions \ref{as:2}--\ref{as:4} below, we construct a nested reduced rank structure on $\C(s,t)$, to fully appreciate both the multivariate and the functional nature of the problem.

\begin{structure}{(Global reduced-rank structure)}\label{as:2}
\begin{align*}
\C_0(s,t) = \U_0\C_0^*(s,t)\V_0\trans, \qquad s\in \mathcal{S}, t\in \mathcal{T},
\end{align*}
where $\U_0\in \mathbb{R}^{d\times r_y}$ with $r_y\leq d$, $\V_0\in \mathbb{R}^{p\times r_x}$ with $r_x\leq p$, and $\C_0^*(s,t)$ is an $r_y\times r_x$ latent regression surface. Without loss of generality, we assume $\U_0\trans\U_0=\I_{r_y}$ and $\V_0\trans\V_0=\I_{r_x}$.
\end{structure}

In Structure \ref{as:2}, $\U_0$ and $\V_0$ are designed to capture the ``global'' effects of the functional association, i.e., it implies that the association between $\y(t)$ and $\x(t)$ is driving by some lower-dimensional latent functional responses and latent predictors that are formed as some linear combinations of the original functional responses and predictors, respectively. That is, it implies that
\begin{align*}
\y^*(t) = \int_{\mathcal{S}}\C_0^*(s,t)\x^*(s)ds + \beps^*(t),
%\y^*(t) = \A^*(t)\left\{\int\B^{*\trans}(s)\x^*(s)ds\right\}+ \beps^*(t),
\end{align*}
where $\y^*(t)=\U_0\trans\y(t)$, $\x^*(s)=\V_0\trans\x(s)$ and $\beps^*(t)=\U_0\trans\beps (t)$.
%$\y^*(t) = \mathcal{P}_{\U_0}\y(t)$, $\x^*(s)= \mathcal{P}_{\V_0}\x(s)$ and $\beps^*(t) =\mathcal{P}_{\U_0}\beps(t)$, with $\mathcal{P}_{\U_0} = \U_0(\U_0\trans\U_0)^{-1}\U_0\trans$ and $\mathcal{P}_{\V_0} =\V_0(\V_0\trans\V_0)^{-1}\V_0\trans$ being the unique projection matrices onto to subspaces spanned by $\U_0$ and $\V_0$, respectively.
%Dimension reduction is achieved when $r_y\leq d$ and/or $r_x\leq p$. 
When $r_y < d$ and/or $r_x < p$, our model achieves great dimensionality reduction and parsimony while retaining flexibility. It includes the structures: $\C_0(s,t)$, $\U_0\C_0^*(s,t)$ and $\C_0^*(s,t)\V_0\trans$ as special cases. This structure is particularly helpful for simultaneously modeling a large number of functional responses and predictors that are highly correlated across $s$ or $t$.

It is conventional to take a basis expansion and truncation approach to facilitate the modeling of the latent regression surface $\C_0^*(s,t)$ \citep{ramsay2005}, for inducing its smoothness over both $s$ and $t$ and converting the infinite dimensional problem to finite dimensional. Specifically, we represent the latent regression surface $\C_0^*(s,t)$ as
%\begin{structure}\label{as:3}
\begin{align}
\C_0^*(s,t) \approx (\I_{r_y}\otimes \bPsi\trans(t))\C_0^*(\I_{r_x}\otimes \bPhi(s)), \qquad \C_0^* \in \mathbb{R}^{(J_yr_y)\times (J_xr_x)},\label{eq:basisexpansion}
\end{align}
where $\I_{a}$ denotes the $a \times a$ identity matrix, $\bPhi(s)= [\phi _1(s),\ldots, \phi_{J_x}(s)]\trans$ consists of a set of basis functions with $\J_{\phi\phi}=\int_{\mathcal{S}} \bPhi(s)\bPhi\trans(s)ds$ being positive definite (p.d.), and similarly $\bPsi(t)= [\psi _1(t),\ldots, \psi_{J_y}(t)]\trans$ with $\J_{\psi\psi}=\int_{\mathcal{T}}\bPsi(t)\bPsi\trans(t)dt$ being p.d. %and $\C_0^*$ is an $(J_yr_y)\times (J_xr_x)$ coefficient matrix.
%Here the smooth functional entries of $\C^*(s,t)$ are expanded using the two sets of basis functions $\bPsi(t)$ and $\bPhi(s)$, which are also used to expand $\y(t)$ and $\x(s)$.
%\end{structure}
Here we assume the basis functions are given, such as spline, wavelet, and Fourier basis; also, for simplicity, we have assumed all the responses or the predictors share the same set of basis, either $\bPsi(t)$ or $\bPhi(s)$, respectively. An alternative is to take a functional principal component analysis (FPCA) or functional canonical correlation analysis (FCCA), in which the basis are obtained as eigenfunctions of covariance operators of $\y(t)$ and $\x(s)$. While with any given number of components such a data-driven basis expansion can explain most of the variation in the $\ell_2$ sense, the analysis is much more complicated as it then involves the estimation of the unknown basis. We thus take the basis as chosen with a sufficiently large number of components and invoke regularization in model estimation.

With the expansion in \eqref{eq:basisexpansion}, it boils down to consider the modeling of the high-dimensional coefficient matrix $\C_0^*$. We further explore a potential low-rank structure in $\C_0^*$.
\begin{structure}{(Local reduced-rank structure)}\label{as:4}
\begin{align*}
\mbox{rank}(\C_0^*)\leq r,
\end{align*}
for $r\leq \min(J_yr_y,J_xr_x)$; that is, $\C_0^* = \A_0^*\B_0^{*\trans}$ for some $\A_0^*\in \mathbb{R}^{(J_yr_y)\times r}$, $\B_0^*\in \mathbb{R}^{(J_xr_x)\times r}$.
\end{structure}
As this structure induces the dependency between the latent responses and the latent predictors through their basis-expanded representations, we achieve a finer dimension reduction at the ``local'' level.

The approximation error in \eqref{eq:basisexpansion} can be controlled under reasonable conditions. Assume that the $\left\lfloor \gamma \right\rfloor $th order derivative of
each function in $\mathbf{C}_{0}^{\ast }(s,t)$ satisfies the H\"{o}lder
condition of order $\gamma -\left\lfloor \gamma \right\rfloor $ with $\gamma
>1/2$, where $\left\lfloor \gamma \right\rfloor $ is the biggest integer
strictly smaller than $\gamma $. This smoothness condition together with
Structures \ref{as:2}--\ref{as:4} imply that the regression surface $\C_0(s,t)$  approximately admits a \textit{nested reduced-rank
representation},
\begin{align}
\sup_{s\in \mathcal{S},t\in \mathcal{T}}|\C_0(s,t)- \U_0(\I_{r_y}\otimes \bPsi\trans(t))\A_0^*\B_0^{*\trans}(\I_{r_x}\otimes \bPhi(s))\V_0\trans|=O(J_{y}^{-\gamma
}+J_{x}^{-\gamma }).\label{eq:expansion}
\end{align}
We can choose the number of basis functions satisfying $%
J_{y}\rightarrow \infty $ and $J_{x}\rightarrow \infty $ as $n\rightarrow
\infty $, so that the above approximation error vanishes. Indeed, this is allowed in our non-asymptotic theoretical analysis which provides a high-probability prediction error bound; see Section \ref{theory} for details.

Model \eqref{eq:model} then becomes
\begin{align}
%\y(t) = & \int \C(s,t)\x(s) ds + \beps(t)\\
\y(t) & \approx \int_{\mathcal{S}} \U_0(\I_{r_y}\otimes \bPsi\trans(t))\A_0^*\B_0^{*\trans}(\I_{r_x}\otimes \bPhi(s))\V_0\trans\x(s) ds + \beps(t)\notag\\
% & \approx \U_0(\I_{r_y}\otimes \bPsi\trans(t))\A_0^*\B_0^{*\trans}\left\{\int_{\mathcal{S}}(\I_{r_x}\otimes \bPhi(s))\V_0\trans \x(s)ds\right\} + \beps(t)\notag\\
 & \approx (\I_{d}\otimes \bPsi\trans(t))(\U_0\otimes \I_{J_y})\A_0^*\B_0^{*\trans}(\V_0\trans\otimes \I_{J_x})\left\{\int_{\mathcal{S}}(\I_{p}\otimes \bPhi(s)) \x(s)ds\right\} + \beps(t).\label{eq:model2}
%=& \U(\I_{r_y}\otimes \bPsi\trans(t))\A^*\{\int\B^{*\trans}(\I_{r_x}\otimes\bPhi(s))\V\trans\x(s)ds\}+ \beps(t)\\
% & = \U\A^*(t)\left\{\int\B^{*\trans}(s)\V\trans\x(s)ds\right\}+ \beps(t),\label{eq:model2}
\end{align}
We remark that $\U_0$, $\V_0$, $\A_0^*$ and $\B_0^*$ are not fully identifiable individually up to rotation or nonsingular transformation, similar to the settings in conventional reduced-rank estimation; nevertheless, the structure as a whole is well-defined and identifiable.

It is worthwhile to mention a few special cases. When the low-dimensional structures do not present at all, i.e., $r_x=p$, $r_y=d$ and $r=\min(J_xr_x,J_yr_y)$, the model becomes $\C_0(s,t)=(\I_{d}\otimes \bPsi\trans(t))\C_0^*(\I_{p}\otimes \bPhi(s))$, for which the least squares estimation is equivalent to separately regressing each response $y_k(t)$ on $\x(s)$ and hence there is no gain of conducting multivariate analysis. When the global structure does not present, i.e., $r_x=p$ and $r_y=d$, the model reduces to a reduced-rank functional model as in \citet{Ebaid2008}. %In our new formulation, the design of $\U_0$ and $\V_0$ shares similar spirit with the envelope models \citep{cook2013} proposed in the context of multivariate regression.

% Due to the global low-rank structure, the model implies that
% \begin{align*}
% \y^*(t) = (\I_{r_y}\otimes \bPsi\trans(t))\A_0^*\B_0^{*\trans}\left\{\int_{\mathcal{S}}(\I_{r_x}\otimes \bPhi(s))\x^*(s)ds\right\} + \beps^*(t),
% %\y^*(t) = \A^*(t)\left\{\int\B^{*\trans}(s)\x^*(s)ds\right\}+ \beps^*(t),
% \end{align*}
% where $\y^*(t) = \U_0\trans\y(t)$, $\x^*(s)= \V_0\trans\x(s)$ and $\beps^*(t) = \U_0\trans\beps(t)$.

% It is also helpful to write out the model for each $y_k(t)$, to better see the parameter sharing induced by the nested low-rank structure,
% \begin{align*}
% y_k(t) & = \sum_{l=1}^{p}\sum_{i=1}^{r_y}\sum_{j=1}^{r_x}a_{ki}b_{lj}\{\int_{\mathcal{S}}\bPsi\trans(t)\C^*_{ij}\bPhi(s)x_l(s)ds\} + \epsilon_k(t), \qquad k = 1,\ldots, d,
%  %& = \sum_{l=1}^{p}\sum_{i=1}^{r_y}\sum_{j=1}^{r_x}a_{ki}b_{lj}\{\bPsi\trans(t)\C^*_{ij}\J_{\phi\phi}\x_l\} + \epsilon_k(t).
% \end{align*}
% where $a_{ki}$ is the $(k,i)$th entry of $\U_0$, $b_{lj}$ is the $(l,j)$th entry of $\V_0$, and $\C^*_{ij}\in \mathbb{R}^{J_y\times J_x}$ is the $(i,j)$th submatrix of $\C_0^*$.

\subsection{Estimation}

The model estimation at the population level can be conducted through minimizing the mean integrated squared error (MISE) with respect to $\C(s,t)$,
\begin{align}
\mathbb{E}\int_{\mathcal{T}} \left\| \y(t)-  \int_{\mathcal{S}} \C(s,t)\x(s) ds\right\|^2 dt,\label{eq:mise}
\end{align}
where $\|\a\|=\sqrt{\a\trans\a}$ denotes the $\ell_2$ norm. Define the integrated predictor and the integrated response as
\begin{align}
\x =& \int_{\mathcal{S}} (\I_p\otimes \bPhi(s))\x(s)ds, \quad
\y = (\I_d\otimes\J_{\psi\psi}^{-\frac{1}{2}})\int_{\mathcal{T}} (\I_d\otimes \bPsi(t))\y(t)dt,\label{eq:y}
\end{align}
and write
$$
\y(t) = (\I_d\otimes \bPsi\trans(t))(\I_d\otimes\J_{\psi\psi}^{-\frac{1}{2}})\y + (\I_d\otimes \bPsi_{\bot}\trans(t))(\I_d\otimes\J_{\psi_{\bot}\psi_{\bot}}^{-\frac{1}{2}})\y_{\bot},
$$
where $\x\in \mathbb{R}^{J_xp}$, $\y\in\mathbb{R}^{J_yd}$, $\y_{\bot}\in\mathbb{R}^{J_yd}$, and $\int \bPsi(t)\bPsi_{\bot}\trans(t)dt=0$. Under the nested reduced-rank model in \eqref{eq:model2}, the MISE in \eqref{eq:mise} becomes
\begin{align*}
& \mathbb{E}\int_{\mathcal{T}} \left\| \y(t)- (\I_d\otimes \bPsi\trans(t))(\U\otimes\I_{J_y})\A^*\B^{*\trans}(\V\trans\otimes\I_{J_x})\x   \right\|^2 dt\\
= & \mathbb{E}\int_{\mathcal{T}} \left\| (\I_d\otimes \bPsi\trans(t))(\I_d\otimes\J_{\psi\psi}^{-\frac{1}{2}})\y- (\I_d\otimes \bPsi\trans(t))(\U\otimes\I_{J_y})\A^*\B^{*\trans}(\V\trans\otimes\I_{J_x})\x   \right\|^2 dt +\mbox{const}.
%= & \mathbb{E}\left\| (\I_d\otimes\J_{\psi\psi}^{-\frac{1}{2}})\y- (\U\otimes\I_{J_y})\A^*\B^{*\trans}(\V\trans\otimes\I_{J_x})\x\right\|^2_{\W}+\mbox{const},%\label{eq:popcre}
\end{align*}
%where $\W=\I_d \otimes \int_{\mathcal{T}} \bPsi(t)\bPsi\trans(t)dt = \I_d\otimes\J_{\psi\psi}$, with $\|\a\|_{\W}=\sqrt{\a\trans\W\a}$ for any vector $\a$.
As a result, the estimation criterion becomes
% \begin{align}
% \min_{\U,\V,\A^*,\B^*}\left\{E\left\| (\I_d\otimes\J_{\psi\psi}^{-\frac{1}{2}})\y- (\U\otimes\I_{J_y})\A^*\B^{*\trans}(\V\trans\otimes\I_{J_x})\x\right\|^2_{\W}\right\}.
% \end{align}
% or
\begin{align}
\min_{\U,\V,\A^*,\B^*}\left\{\mathbb{E}\left\| \y- (\I_d\otimes\J_{\psi\psi}^{\frac{1}{2}})(\U\otimes\I_{J_y})\A^*\B^{*\trans}(\V\trans\otimes\I_{J_x})\x\right\|^2\right\}.\label{eq:popcre}
\end{align}
This is a generalization of the reduced-rank regression criterion \citep{reinsel1998}. Unlike the latter, however, \eqref{eq:popcre} does not lead to an explicit analytic expression in general.

%From multivariate regression point of view, the matrices $\U$ and $\V$ aim to capture the potential shared structures among certain sub-matrices or blocks of the entire $J_yd\times J_xp$ coefficient matrix of rank $r$, and by doing so it achieves further dimension reduction. With any fixed $\U$ and $\V$, minimizing (\ref{eq:popcre}) with respect to $(\A^*,\B^*)$ becomes a conventional reduced-rank regression problem.

We now consider the corresponding sample estimation problem. Suppose the functional responses and predictors are fully observed for $n$ random subjects over their respective domains, i.e., $(\y_i(t), \x_i(s))$ for $t\in \mathcal{T}$, $s\in\mathcal{S}$, and $i=1,\ldots, n$. The integrated predictors and responses for each subject $i$ can then be computed according to \eqref{eq:y},
\begin{align*}
x_{ilj} =& \int_{\mathcal{S}} \phi_j(s)x_{li}(s)ds,\qquad l=1,\ldots,p; j=1,\ldots,J_x,\\
y_{ikj}^0 =& \int_{\mathcal{T}} \psi_j(t)y_{ki}(t)dt,\qquad k=1,\ldots,d;j=1,\ldots,J_y, \\
y_{ikj}=&\J_{\psi\psi[j,\cdot]}^{-\frac{1}{2}}(y_{ik1}^0, \ldots, y_{ikJ_y}^0)\trans,\qquad k=1,\ldots,d;j=1,\ldots,J_y,
\end{align*}
where $\J_{\psi\psi[j,\cdot]}^{-\frac{1}{2}}$ denotes the $j$-th row of $\J_{\psi\psi}^{-\frac{1}{2}}$. Define $\Y_{\cdot j}= (y_{ikj})_{n\times d}$, for $j=1,\ldots,J_y$, and let $\Y = (\Y_{\cdot1},\ldots, \Y_{\cdot J_y})$. Similarly, define $\X_{\cdot j}= (x_{ilj})_{n\times p}$, and let $\X = (\X_{\cdot1},\ldots, \X_{\cdot J_x})$. We write $\A^*=(\A_{1\cdot}\trans,\ldots,\A_{r_y\cdot}\trans)\trans$ where $\A_{h\cdot}\in \mathbb{R}^{J_y\times r}$ for $h=1,\ldots, r_y$, and $\B^*=(\B_{1\cdot}\trans,\ldots,\B_{r_x\cdot}\trans)\trans$ where $\B_{h\cdot}\in \mathbb{R}^{J_x\times r}$ for $h=1,\ldots, r_x$. Define $\widetilde{\A}_{h\cdot} = \J_{\psi\psi}^{\frac{1}{2}}\A_{h\cdot}$ and $\widetilde{\A}^*=(\widetilde{\A}_{1\cdot}\trans,\ldots,\widetilde{\A}_{r_y\cdot}\trans)\trans$. Since $\J_{\psi\psi}$ is nonsingular, it suffices to consider the estimation of $\widetilde{\A}^{*}$ instead of $\A^{*}$. It is necessary to  rearrange the rows of $\widetilde{\A}^*$ and $\B^*$, i.e., let $\A = (\A_{\cdot1}\trans,\ldots,\A_{\cdot J_y}\trans)\trans$ where $\A_{\cdot j} \in \mathbb{R}^{r_y\times r}$ is formed by collecting the $j$th row of each $\widetilde{\A}_{h\cdot}$, and $\B = (\B_{\cdot1}\trans,\ldots,\B_{\cdot J_x}\trans)\trans$ where $\B_{\cdot j} \in \mathbb{R}^{r_x\times r}$ is formed by collecting the $j$th row of each $\B_{h\cdot}$. Finally, these matrix notations allow us to write the sample MISE criterion as a nested reduced-rank regression problem,
\begin{align}
\min_{\C}\|\Y-\X\C\|_{\rm F}^2,\qquad s.t.\, \C=(\I_{J_x} \otimes \V)\B\A\trans(\I_{J_y}\otimes \U\trans).\label{eq:samcre}
\end{align}
%It is now clear that 
Thus from matrix approximation point of view, $\U$ and $\V$ are designed to capture the shared column and row spaces among the blockwise sub-matrices of $\C$, which, as a whole, is also of low rank. Figure \eqref{fig:nrr} shows a conceptual diagram of this nested reduced-rank structure.

%Third diagram
% Define the layers to draw the diagram
\pgfdeclarelayer{background}
\pgfdeclarelayer{foreground}
\pgfsetlayers{background,main,foreground}
% Define block styles used later
\tikzstyle{xu}=[draw, fill=blue!20, text width=1em,
    text centered, minimum height=8em,drop shadow]
\tikzstyle{xd}=[draw, fill=green!20, text width=1em,
    text centered, text height=1em,drop shadow]
\tikzstyle{xv}=[draw, fill=red!20, text width=4em,
    text centered, minimum height=1em,drop shadow]
\tikzstyle{ann} = [text width=5em, text centered]
\tikzstyle{by} = [xu, text width=6em, fill=red!20,
    minimum height=16em, rounded corners, drop shadow]
\tikzstyle{xulong}=[draw, fill=blue!20, text width=2em,
    text centered, minimum height=19.4em,drop shadow]
\tikzstyle{xvlong}=[draw, fill=red!20, text width=16em,
    text centered, minimum height=2em,drop shadow]
\tikzstyle{bx} = [xu, text width=6em, fill=blue!20,
    minimum height=10em, rounded corners, drop shadow]
\tikzstyle{myarrows}=[line width=0.5mm,draw=black,-triangle 45,postaction={draw, line width=1.5mm, shorten >=4mm, -}]
\usetikzlibrary{arrows, decorations.markings}
\tikzstyle{vecArrow} = [thick, decoration={markings,mark=at position
   1 with {\arrow[semithick]{open triangle 60}}},
   double distance=1.4pt, shorten >= 5.5pt,
   preaction = {decorate},
   postaction = {draw,line width=1.4pt, white,shorten >= 4.5pt}]
\tikzstyle{innerWhite} = [semithick, white, line width=1.4pt, shorten >= 4.5pt]
\tikzstyle{vecArrow2} = [thick, decoration={markings,mark=at position
   0.75 with {\arrow[semithick]{open triangle 60}}},
 double distance=1.4pt, shorten >= 17.85pt,
   preaction = {decorate},
   postaction = {draw,line width=1.4pt, white,shorten >= 17.85pt}]
\tikzstyle{innerWhite2} = [semithick, white, line width=1.4pt, shorten >= 18pt]
\usetikzlibrary{arrows,positioning}
\tikzset{
    %Define standard arrow tip
    >=stealth',
    %Define style for boxes
    punkt/.style={
           rectangle,
           rounded corners,
           draw=black, very thick,
           text width=6.5em,
           minimum height=2em,
           text centered},
    % Define arrow style
    pil/.style={
           ->,
           thick,
           shorten <=2pt,
           shorten >=2pt,}
}
% Define distances for bordering
\def\blockdist{2}
\def\blockuv{4em}
\def\edgedist{2}
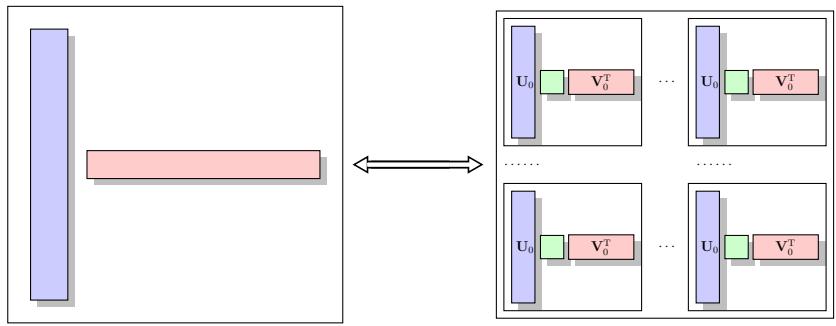
\begin{figure}[htp]
\centering
\begin{tikzpicture}[remember picture,scale = 0.45,every node/.style={scale=0.45}]

    \node at (-2.5,0.5) (xulong) [xulong]{};
    \path (xulong.east) + (4,0) node (xvlong) [xvlong]{};
    \node [draw=black, fit= (xulong) (xvlong),inner sep=0.3cm,scale=1/0.45] {};

    \path (xvlong.east)+(6,0) node (dots2) [ann] {$\cdots\cdots$};
    \path (dots2.north)+(0,2.2) node (xu1) [xu] {$\U_0$};
    \path (xu1.east)+(0.5,0) node (xd1) [xd] {};
    \path (xd1.east)+(1.1,0) node (xv1) [xv] {$\V_0\trans$};
    \path (dots2.south)+(0,-2.2) node (xuk) [xu] {$\U_0$};
    \path (xuk.east)+(0.5,0) node (xdk) [xd] {};
    \path (xdk.east)+(1.1,0) node (xvk) [xv] {$\V_0\trans$};

    \draw [vecArrow](xvlong.east)+(1.5,0)--(dots2.west);
    \draw [innerWhite](xvlong.east)+(1.5,0)--(dots2.west);

    \draw [vecArrow2](dots2.west)+(-1,0)--(xvlong.east);
    \draw [innerWhite2](dots2.west)+(-1,0)--(xvlong.east);

    \path (xv1.east) + (1,0) node (dots4) [ann] {$\cdots$};
    \path (dots4.east) + (0,0) node (xu3) [xu] {$\U_0$};
    \path (xu3.east)+(0.5,0) node (xd3) [xd] {};
    \path (xd3.east)+(1.1,0) node (xv3) [xv] {$\V_0\trans$};
    \path (dots2.east)+ (4.5,0) node (dots3) [ann] {$\cdots\cdots$};
    \path (xvk.east) + (1,0) node (dots5) [ann] {$\cdots$};
    \path (dots5.east) + (0,0) node (xu4) [xu] {$\U_0$};
    \path (xu4.east)+(0.5,0) node (xd4) [xd] {};
    \path (xd4.east)+(1.1,0) node (xv4) [xv] {$\V_0\trans$};

 \node [draw=black, fit= (xu1) (xv1),inner sep=0.1cm,scale=1/0.45] {};
 \node [draw=black, fit= (xuk) (xvk),inner sep=0.1cm,scale=1/0.45] {};
 \node [draw=black, fit= (xu3) (xv3),inner sep=0.1cm,scale=1/0.45] {};
 \node [draw=black, fit= (xu4) (xv4),inner sep=0.1cm,scale=1/0.45] {};
 \node [draw=black, fit= (xu1) (xu4) (xv4),inner sep=0.2cm,scale=1/0.45] {};

\end{tikzpicture}
\caption{A diagram of the nested reduced-rank matrix representation.}\label{fig:nrr}
%\vspace{-0.35cm}
\end{figure}

Thus far the functional responses and predictors are treated as given. In practical situations, however, the functional data are often observed not continuously or densely, but at discrete points. It is certainly preferable to account for this uncertainty in statistical analysis, but we do not pursue this complication in the current work. Following \citet{ramsay2005}, the preceding integrals are approximated by finite Riemann sums with discrete observations. Suppose for $i=1,\ldots,n$, we observe
$\y_{i}\left( t\right)=(y_{1i}\left( t\right),\ldots,y_{di}\left( t\right))\trans$ at discretized time points $t_{i,v}$, for $v=1\ldots,m_{i}$, and
$\x_{i}\left( s\right)=(x_{1i}\left( s\right),\ldots, x_{pi}\left( s\right))\trans$
at discretized time points $s_{i,u}$, for $u=1,\ldots,g_{i}$. Based on (\ref{eq:y}), we compute
\begin{align*}
x_{ilj} =& \sum_{u=2}^{g_i} \phi_j(s_{i,u})x_{li}(s_{i,u})(s_{i,u}-s_{i,u-1}),\qquad j=1,\ldots,J_x,\\
y_{ikj}^0 =& \sum_{v=2}^{m_i} \psi_j(t_{i,v})y_{ki}(t_{i,v})(t_{i,v}-t_{i,v-1}),\qquad j=1,\ldots,J_y.
%y_{ikj}=&\J_{\psi\psi}^{-\frac{1}{2}}(y_{ik1}^0, \ldots, y_{ikJ_y}^0)\trans,\qquad j=1,\ldots,J_y.
\end{align*}

\subsection{Other Applications}

The applicability of the nested reduced-rank estimation is beyond the functional setup. An interesting application is in high-dimensional vector autoregressive (VAR) modeling in multivariate time series analysis. Let $\y_t \in \mathbb{R}^p$ be the observed multivariate time series at time $t$. Consider a VAR model of order $h$,
\begin{align*}
	\y_t=\A_1 \y_{t-1} + \ldots + \A_h \y_{t-h} + \e_t = \A \x_{t-1} + \e_t,\qquad t=1,\ldots, T,
\end{align*}
where $\A_i \in \mathbb{R}^{p \times p}$, $\A=(\A_1,\ldots,\A_{h}) \in \mathbb{R}^{p \times hp}$, $\x_{t-1}=(\y_{t-1}\trans, \ldots, \y_{t-h}\trans)\trans \in \mathbb{R}^{hp}$, and $\e_t \in \mathbb{R}^{p}$ is a zero-mean innovative process. Stationary reduced-rank VAR model was introduced in \cite{luetkepohl1991}, where the coefficient matrix $\A$ is assumed to be of low rank. In high-dimensional scenarios, it is possible that (1) some linear combinations of the multivariate time series $\y_t$ are processes of pure noise, and (2) the dynamics of $\y_t$ is driven by its lags only through some linear combinations. This gives arise a nested reduced-rank structure. Specifically, the global structure can be modeled as
$$
\A_i = \U_0\A_i^*\V_0\trans, \qquad i = 1,\ldots,h,
$$
where $\U_0\in \mathbb{R}^{p\times r_1}$ with $r_1\leq p$, $\V_0\in \mathbb{R}^{p\times r_2}$ with $r_2\leq p$, satisfying $\U_0\trans\U_0=\I_{r_1}$ and $\V_0\trans\V_0=\I_{r_2}$. The local low-dimensional structure can be modeled by letting the matrix $(\A_1^*,\ldots,\A_h^*) \in \mathbb{R}^{r_1\times(hr_2)}$ be of low rank. As such, $\V_0\trans\y_t$ gives the latent principal time series, and $\U_0^{\perp\trans}\y_t$ are pure noise where $\U_0^\perp \in \mathbb{R}^{p\times (p-r_1)}$ and $\U_0\trans\U_0^\perp = \0$.

Another potential application is in surveillance video processing. In recent years, the sparse plus low-rank decomposition has been a popular method for surveillance video decoding, in which the low-rank component represents the background and the sparse component captures the moving objects. Since the surveillance video frames are usually with a static or gradually changed background, using a nested reduced-rank component with an extra global reduction scheme may improve the efficiency of background  representation by dramatically reducing the temporal redundancy. These ideas will be further explored in our future work.

\section{Computation}\label{computation}
%Both the global and local features of $\C(s,t)$ can then be effectively captured. We shall point out that, $\U$, $\V$, $\A^*$ and $\B^*$ are not completely identifiable individually, similar to the settings in conventional reduced rank estimation. Nevertheless, the nested reduced-rank structure itself is completed determined. Without loss of generality, we always assume $\U\trans\U=\I_{r_y}$ and $\V\trans\V=\I_{r_x}$ in the sequel.
\subsection{A Blockwise Coordinate Descent Algorithm}

%Minimizing (\ref{eq:samcre}) does not lead to explicit solution in general. 

%We design an iterative optimization algorithm to solve \eqref{eq:samcre}. As discussed earlier, the matrices $(\U,\V,\A,\B)$ are not fully identifiable in the parameterization $\C=(\I_{J_x} \otimes \V)\B\A\trans(\I_{J_y}\otimes \U\trans)$, thus our interest lies in the estimation of $\C$, i.e., identifying the subspaces uniquely determined by $\U$ and $\V$ and the reduced-rank structure determined by $\B\A\trans$. To facilitate stable computation, we always set $\U\trans\U=\I_{r_y}$, $\V\trans\V=\I_{r_x}$.

When $\U$ and $\V$ are held fixed, minimizing (\ref{eq:samcre}) becomes a reduced-rank regression \citep{reinsel1998},
\begin{align}
\min_{\A,\B}\|\Y_L-\X_L\B\A\trans\|_{\rm F}^2,\label{eq:solveAB}
%\min\|\Y^{(L)}-\X^{(L)}\B\A\trans\|_{\rm F}^2
\end{align}
where $\Y_L = \Y(\I_{J_y}\otimes \U)$ and $\X_L=\X(\I_{J_x} \otimes \V)$. One set of explicit solution is given by $\B = (\X_L\trans\X_L)^{-}\X_L\trans\Y_L\V_L(r)$ and $\A = \V_L(r)$, where $\V_L(r)$ consists of the first $r$ eigenvectors of the matrix $\Y_L\trans\X_L(\X_L\trans\X_L)^{-}\X_L\trans\Y_L$. %Note that although $(\A,\B)$ are only identifiable up to nonsingular transformation, but there product $\A\B\trans$ is always fully identifiable.
%$\widetilde{\C}=(\X_L\trans\X_L)^{-}\X_L\trans\Y_L$ is the least squares estimator, and

%Note that $(\A,\B)$ are identifiable up to nonsingular transformation; %so without loss of generality, we have assumed $\A\trans\A=\I$.

For fixed $\A$, $\B$ and $\V$, the problem becomes
$$
\min_{\U}\sum_{j=1}^{J_y}\|\Y_{\cdot j}-\X_{A,j}\U\trans\|_{\rm F}^2,\qquad s.t.\,\, \U\trans\U=\I_{r_y},
$$
where $\X_{A,j}=\X(\I_{J_x} \otimes \V)\B\A_{\cdot j}\trans$, $j=1,\ldots,J_y$. This is equivalent to
%$$
%\min_{\U} \|\I_{r_y} - (\sum_{j=1}^{J_y}\X\trans_{A,j}\Y_{\cdot j})\U\|_{\rm F}^2,\qquad s.t.\,\, \U\trans\U=\I_{r_y},
%$$
\begin{align}
\min_{\U} \|\I_{d} - (\sum_{j=1}^{J_y}\Y_{\cdot j}\trans\X_{A,j})\U\trans\|_{\rm F}^2,\qquad s.t.\,\, \U\trans\U=\I_{r_y},\label{eq:solveU}
\end{align}
which can be recognized as an orthogonal Procrustes problem and admits an explicit solution $\U= \U_A\V_A\trans$, where $\U_A\D_A\V_A\trans$ is the SVD of the matrix $\sum_{j=1}^{J_y}\Y_{\cdot j}\trans\X_{A,j}$.

In order to update $\V$, we consider fix $\A$ and $\U$ and write the problem with respect to both $\B$ and $\V$ as
\begin{align}
%\|\Y(\I_{J_y}\otimes \U)\A-\sum_{j=1}^{J_x}\X_{\cdot j}\V\B_{\cdot j}\|_{\rm F}^2 + \mbox{const}\\
\min_{(\B,\V)} \|\y_B-\X_B\mbox{vec}(\V)\|^2,\qquad s.t.\,\,\V\trans\V=\I_{r_x},\label{eq:solveBV}
\end{align}
where $\y_B = \mbox{vec}\{\Y(\I_{J_y}\otimes \U)\A\}$ and $\X_B = \sum_{j=1}^{J_x}(\B_{\cdot j}\trans\otimes \X_{\cdot j})$. Here $\mbox{vec}(\cdot)$ is the vectorization operator for converting a matrix to a vector by concatenating its columns; we will also use $\mbox{dvec}(\cdot)$ to denote the corresponding de-vectorization operation. It is not necessary to solve (\ref{eq:solveBV}) fully, as long as the updates can decrease the value of the original objective function in \eqref{eq:samcre} (which is the same as in \eqref{eq:solveBV} for fixed $\A$ and $\U$). We thus propose the following one-step update of $(\B,\V)$. Ignoring the orthogonality constraints on $\V$ for a moment, we first compute the least squares solution $\widetilde{\V}=\mbox{dvec}\{ (\X_B\trans\X_B)^{-1}\X_B\trans\y_B\}$. We then perform QR decomposition of $\widetilde{\V}$, i.e., $\widetilde{\V}=\Q_B\R_B$, and let $\V=\Q_B$ and update $\B_{\cdot j}$ as $\R_B\B_{\cdot j}$. This step ensures the orthogonality of $\V$, and makes the objective function decrease. 

Algorithm \ref{alg:NRRR} presents the proposed algorithm, for any fixed triplets of rank values $(r,r_x,r_y)$. The matrices $\A$, $\B$, $\U$ and $\V$ are alternatingly updated according to \eqref{eq:solveAB}, \eqref{eq:solveU} and \eqref{eq:solveBV}. The objective in \eqref{eq:samcre} is monotone decreasing along the iterations, and consequently the convergence to a limiting point is guaranteed.

\begin{algorithm}[htp]
\caption{Nested Reduced-Rank Regression}\label{alg:NRRR}
\begin{algorithmic}
\STATE Initialize $\U^0\in \mathbb{R}^{d\times r_y}$, $\V^0\in\mathbb{R}^{p\times r_x}$.
%$\A^{0}\in \mathbb{R}^{J_yr_y\times r}$ and $\B^{0}\in\mathbb{R}^{J_xr_x\times r}$.
\STATE Set $k\gets0$.
\REPEAT
\STATE (1). RRR updates:
%\begin{align*}
%(\A^{k+1},\B^{k+1}) \gets\arg\min_{(\A,\B)}\|\Y_L-\X_L\B\A\trans\|_{\rm F}^2,
%\end{align*}
\begin{align*}
\B^{k+1} \gets (\X_L^{\trans}\X_L)^{-}\X_L\trans\Y_L\V_L(r),\qquad \A^{k+1} \gets \V_L(r),
\end{align*}
where $\Y_L= \Y(\I_{J_y}\otimes \U^{k})$ and $\X_L=\X(\I_{J_x} \otimes \V^{k})$, and $\V_L(r)$ consists of the first $r$ eigenvectors of the matrix $\Y_L\trans\X_L(\X_L\trans\X_L)^{-}\X_L\trans\Y_L$.\\
\vspace{0.2cm}
\STATE (2). Procrustes updates:
\begin{align*}
\U^{k+1}\gets \U_A\V_A\trans,
\end{align*}
where $\U_A\D_A\V_A\trans$ is the SVD of $\sum_{j=1}^{J_y}\Y_{\cdot j}\trans\X_{A,j}$, with $\X_{A,j}=\X(\I_{J_x} \otimes \V^k)\B^{k+1}\A_{\cdot j}^{k+1 \trans}$.
\vspace{0.2cm}
\STATE (3). QR updates:
\begin{align*}
&\V^{k+1}\gets \Q_B,%\\
%&\B_{\cdot j}^{k+1} \gets\R^{(B)}\B_{\cdot j}^{k+1},
\end{align*}
where $\Q_B\R_B$ is the QR decomposition of $\mbox{dvec}\{ (\X_B\trans\X_B)^{-1}\X_B\trans\y_B\}$, with $\y_B = \mbox{vec}\{\Y(\I_{J_y}\otimes \U^{k+1})\A^{k+1}\}$ and $\X_B = \sum_{j=1}^{J_x}(\B_{\cdot j}^{k+1 \trans}\otimes \X_{\cdot j})$.\\
\vspace{0.2cm}
\STATE Set $k\gets k+1$.
\UNTIL{convergence, i.e., $\|\C^{k+1}-\C^{k}\|/\|\C^k\|\leq \epsilon=10^{-4}$, where $\C^{k}=(\I_{J_x} \otimes \V^k)\B^k\A^{k\trans}(\I_{J_y}\otimes \U^{k\trans})$.}
\end{algorithmic}
\end{algorithm}

%\clearpage

\subsection{Initial Estimator and Rank Selection}
%We propose a natural way of obtaining 

Some initial estimates of $(\U,\V)$ are required for running the proposed algorithm for a specified set of rank values $(r, r_x, r_y)$. The coefficient matrix $\C$ in (\ref{eq:samcre}) takes the form $\C=(\I_{J_x} \otimes \V)\B\A\trans(\I_{J_y}\otimes \U\trans)$, which implies that $\mbox{rank}(\C)\leq r$. Therefore, ignoring the global structure in $(\U,\V)$ for a moment, $\C$ can be directly estimated by a conventional reduced-rank regression of $\Y$ on $\X$, i.e.,
\begin{align}
\min_{\C}\|\Y-\X\C\|_{\rm F}^2,\qquad s.t.\,\,\mbox{rank}(\C)\leq r,\label{eq:rrr}
\end{align}
and the minimizer is given by $\widetilde{\C}=\widetilde{\B}\widetilde{\A}\trans$, $\widetilde{\B}=(\X\trans\X)^{-}\X\trans\Y\V(r)$, $\widetilde{\A}=\V(r)$, where $\V(r)$ consists of the first $r$ eigenvectors of the matrix $\Y\trans\X(\X\trans\X)^{-}\X\trans\Y$. The $\widetilde{\B}$ and $\widetilde{\A}$ can be viewed as approximations to $(\I_{J_x} \otimes \V)\B$ and $(\I_{J_y}\otimes \U)\A$, respectively. Therefore, an initial estimator of $\V$ can be obtained from
\begin{align*}
(\V^0,\B^0) = \arg\min_{(\V,\B)}\|\widetilde{\B}-(\I_{J_x} \otimes \V)\B\|_{\rm F}^2.
\end{align*}
Write $\widetilde{\B}=(\widetilde{\B}_1\trans,\ldots,\widetilde{\B}_{J_x}\trans)\trans$ where each $\widetilde{\B}_i\in \mathbb{R}^{p\times r}$, $i=1,\ldots,J_x$. Based on Eckart-Young Theorem, it can be easily shown that
\begin{align*}
\V^0 = \widetilde{\U}_{\widetilde{B}}(r),%\qquad (\B^0_{\cdot1},\ldots,\B_{\cdot J_x}) = \widetilde{\V}_B(r)\widetilde{\D}_B(r),
\end{align*}
where $\widetilde{\U}_{\widetilde{B}}(r)$ consists of the first $r$ left singular vectors of the the matrix $(\widetilde{\B}_1,\ldots,\widetilde{\B}_{J_x})$. Similarly, write $\widetilde{\A}=(\widetilde{\A}_1\trans,\ldots,\widetilde{\A}_{J_y}\trans)\trans$ where each $\widetilde{\A}_i\in \mathbb{R}^{d\times r}$, then an initial estimator of $\U$ is obtained from minimizing $\|\widetilde{\A}-(\I_{J_y} \otimes \U)\A\|_{\rm F}^2$ with respect to $(\U,\A)$, so that
\begin{align*}
\U^0 = \widetilde{\U}_{\widetilde{A}}(r),%\qquad (\B^0_{\cdot1},\ldots,\B_{\cdot J_x}) = \widetilde{\V}_B(r)\widetilde{\D}_B(r),
\end{align*}
where $\widetilde{\U}_{\widetilde{A}}(r)$ consists of the first $r$ left singular vectors of the the matrix $(\widetilde{\A}_1,\ldots,\widetilde{\A}_{J_y})$.

To choose an optimal set of rank values $(r,r_x,r_y)$, the $K$-fold cross validation procedure can be used, which, however, can be quite computationally expensive for large-scale problems. Here we propose to select $(r,r_x,r_y)$ based on a Bayesian Information Criterion (BIC) \citep{schwarz1978}, because of its computational efficiency and promising performance in regularized estimation. Denote $\wh{\C}(r,r_x,r_y)$ as the estimator of $\C$ by solving (\ref{eq:samcre}) with the rank values fixed at some $(r,r_x,r_y)$. We define
\begin{align}
\mbox{BIC}(r,r_x,r_y) = ndJ_y\log\left\{\mbox{SSE}(r,r_x,r_y)/(ndJ_y)\right\} + \log(ndJ_y)df(r,r_x,r_y),\label{eq:bic}
\end{align}
where $\mbox{SSE}(r,r_x,r_y)=\|\Y-\X\wh{\C}(r,r_x,r_y)\|_{\rm F}^2$ stands for the sum of squared errors and $df(r,r_x,r_y)$ is the effective degrees of freedom of the model. We use the number of free model parameters to estimate $df(r,r_x,r_y)$,
\begin{align}
\wh{df}(r,r_x,r_y)=r_x\{r(\X)/J_x-r_x\} + r_y(d-r_y) + (J_y r_y+J_x r_x-r)r.\label{eq:df}
\end{align}
When $r_y=d$, $r_x=r(\X)/J_x$, the above formula gives $\wh{df}(r,r(\X)/J_x,d)=(J_yr_y + r(\X)-r)r$, which is exactly the effective number of parameters in a rank-$r$ reduced-rank regression model \citep{Muhk2013}. The difference in the number of parameters is
$(J_y d - J_y r_y)(r-r_y/J_y) + (r(\X)-J_x r_x)(r-r_x/J_x)$.

With the above BIC criterion, a three-dimensional grid search procedure of the rank values can be performed, and the best model is chosen as the one with the smallest BIC value. On the other hand, note that the global structure of the predictors determined by $r_x$, the global structure of the responses determined by $r_y$, and the local structure determined by $r$ are designed to realize different low-dimensional aspects of $\C$. As such, a one-at-a-time selection approach works well in practice. We first set $r_x=p$, $r_y=d$, and select the best local rank $\wh{r}$ among the models with $1\leq r\leq \min(r(\X),J_yd)$. We then fix the local rank at $\wh{r}$, and repeat the similar procedure to determine $\wh{r}_x$ and $\wh{r}_y$, one at a time. Finally, with fixed $\wh{r}_x$ and $\wh{r}_y$, we refine the estimation of $r$. This approach is adapted in all our numerical studies and works quite well.

\section{Theoretical Analysis}\label{theory}

Our theoretical analysis concerns the fundamental nested reduced-rank regression setup,
$$
\Y = \X\C_0 + \E, \qquad s.t.\,\, \C_0=(\I_{J_x} \otimes \V_0)\B_0\A_0\trans(\I_{J_y}\otimes \U_0\trans).
$$ 
Accordingly, the objective function is defined as 
\begin{align*}
	\Q_n(\V,\B,\A,\U)=\|\Y - \X (\I_{J_x} \otimes \V) \B \A\trans (\I_{J_y} \otimes \U\trans) \|_F^2,
\end{align*}
and the NRRR estimator is obtained as 
\begin{align*}
	(\widehat \V, \widehat \B, \widehat \A, \widehat \U) \in \arg\min_{\V,\B,\A,\U} \Q_n (\V,\B,\A,\U).
\end{align*}
To facilitate the analysis, it is necessary to make the components ($\V_0, \B_0, \A_0, \U_0$) identifiable individually; we defer the discussion until presenting the main results. Here the integrated response and predictor matrices from functional data are treated as given, as the functional approximation aspect of the problem is not our focus. We have assumed that the rank values are known. Even so, the non-convexity of the NRRR problem, induced by the complex nested low-rank matrix decomposition, makes the theoretical analysis challenging. 
%Here $\C_{0} = (\I_{J_x}\otimes \V_0)\B_0\A_0\trans(\I_{J_y} \otimes \U_0\trans)$ is the true parameter matrix from Model \eqref{eq:model}.
%In this section, we conduct both asymptotic and non-asymptotic analysis to demonstrate the superiority of the proposed NRRR estimator. The asymptotic analysis provides the consistency property of the NRRR estimator while the non-asymptotic analysis demonstrates that the performance of the NRRR estimator is at least comparable to the one of the classical reduced-rank regression estimator.
%\subsection{Asymptotic Analysis}

We need the following conditions for our asymptotic analysis.

\begin{assumption}\label{asy:assump1}
	$\X\trans\X/n \xrightarrow{a.s.} \bG$ as $n\rightarrow \infty$, where $\bG$ is a fixed, positive-definite matrix.
\end{assumption}

\begin{assumption}\label{asy:assump2}
	Each row $\e_i$ of $\E$ is independently and identically distributed with $\mathbb{E}(\e_i)=\0$ and $\mbox{cov}(\e_i)=\bSig$, where $\bSig$ is positive-definite. 
\end{assumption}
%Then with these two assumptions we have
%\begin{align*}
%	\Psi_n(R^1,R^2,R^3,R^4) \rightarrow \Psi(R^1,R^2,R^3,R^4)=-\Z\trans \W + \frac{1}{2}\Z\trans(\I_{J_y d} \otimes \Gamma)\Z.
%\end{align*}
%Next we state the consistent property of the NRRR estimator.

\begin{theorem}{(Consistency)}\label{asy:thm1}
Suppose Assumptions \ref{asy:assump1} and \ref{asy:assump2} hold. Then there exists a local minimizer $(\widehat\V,\widehat\B,\widehat\A,\widehat\U)$ of $\Q_n(\V,\B,\A,\U)$ such that $\|\widehat \V - \V_0 \|_F=O_p(n^{-\frac{1}{2}})$, $\|\widehat \B - \B_0 \|_F=O_p(n^{-\frac{1}{2}})$, $\|\widehat \A - \A_0 \|_F=O_p(n^{-\frac{1}{2}})$ and $\|\widehat \U - \U_0 \|_F=O_p(n^{-\frac{1}{2}})$.
\end{theorem}

Theorem \ref{asy:thm1} shows the consistency of the NRRR estimation in estimating the components of the nested low-rank structure, in the sense that there exists a local minimizer that is $\sqrt{n}-$consistent. For non-convex problem, such an asymptotic result is what to be expected \citep{fan2001,chen2012reduced}. While the details of the proof are provided in Appendix A, we briefly outline the main steps here. We first parameterize the coefficient matrix $\C_0$ such that the components in its nested low-rank structure, ($\V_0, \B_0, \A_0, \U_0$), can be identifiable. %This step is necessary due to the non-uniqueness of the low-rank decomposition. 
Then a local neighborhood around the true value $\C_0$ with radius $h$ is constructed, denoted as $\mathcal{N}(\C_0,h)$. We then show that for any given $\epsilon > 0$, 
\begin{align*}
	\mathbb{P}\bigg\{ \inf_{\|\check \R^1\|_F=\|\check \R^2\|_F=\|\check \R^3\|_F=\|\check \R^4\|_F=h} & \Q_n(\V_0 + \frac{1}{\sqrt{n}} \R^1, \B_0 + \frac{1}{\sqrt{n}} \R^2, \A_0+ \frac{1}{\sqrt{n}} \R^3, \U_0+ \frac{1}{\sqrt{n}} \R^4) \bigg. \nonumber\\ 
	&  \bigg. > \Q_n(\V_0,\B_0,\A_0,\U_0)  \bigg\} \geq 1-\epsilon,
\end{align*}
with a large enough constant $h$. Here the infimum is taken over the perturbation matrices $\R^1,\R^2,\R^3,\R^4$ (one-to-one transformations of $\check \R^1,\check \R^2,\check \R^3,\check\R^4$) of $\V_0, \B_0, \A_0, \U_0$, respectively, with a fixed Frobenius norm $h$. That is, the objective function evaluated at any boundary point of the neighborhood of radius $h$ is larger than that evaluated at the true value, with arbitrarily large probability. It thus follows that a local minimizer must exist within the neighborhood with a $\sqrt{n}$ convergence rate. 

%This claim states that within $\mathcal{N}(\C_0,h)$, with probability at least $1-\epsilon$, any perturbation on the true coefficient matrix will lead to an increase on the objective function.

%The verification can be done with the help of the identifiability condition imposed on the nested reduced-rank structured true coefficient matrix. For details please see section 1 of the supplemental materials.

%\subsection{Non-Asymptotic Analysis}

We also attempt non-asymptotic analysis, to understand better the behavior of %the proposed method 
NRRR estimator in high-dimensional setups. Let's express the true functional regression surface as $\C_0(s,t) =\{ \I_{d} \otimes {\boldsymbol{\Psi}}(t)\trans\}\{ \I_{d} \otimes \J_{\psi\psi}^{-\frac{1}{2}}\}\widetilde\C_0\trans\{ \I_{p} \otimes {%
\boldsymbol{\Phi }}(s)\} $, where $\widetilde \C_0$ is obtained by a  rearrangement of the columns and rows of $\C_0$. Let $\widehat \C = (\I_{J_{x}} \otimes \widehat \V)\widehat \B\widehat \A\trans(\I_{J_y} \otimes \widehat \U\trans)$ be the NRRR estimator of $\C_0$, and $\widehat{\C}(s,t)$ is obtained by plugging in the corresponding components. 

%{\color{red} (We never define $\E$!)}
% \begin{assumption}[Error distribution]
% \label{as:A1} The random error matrix $\E$ has independent $N \left( 0,\sigma
% ^{2}\right) $ entries. %The eigenvalues of $\left( 1/n\right) \X\trans \X$ are bounded away from $0$ and $\infty$.
% \end{assumption}

% \begin{assumption}
%\label{as:A2} 
%\end{assumption}

% We allow $d \rightarrow \infty $, $p\rightarrow \infty $, $
% r \rightarrow \infty $, $J_y \rightarrow \infty $ and $J_x \rightarrow
% \infty $ as $n \rightarrow \infty $. Let $J_n=q J_y+p J_x$, $J_n r n^{-1}=o\left( 1\right) $ and $r \leq \min \left(r_y J_y, r_x J_x \right) $.

\begin{theorem}\label{THM1}
Suppose the random error matrix $\E$ has independent $N \left( 0,\sigma
  ^{2}\right) $ entries. With probability at least $1-\exp\left\{-\theta^2(r(\X)+ dJ_y)/2\right\}$, we have 
\begin{align*}
\| \X\widehat \C-\X\C_0 \| _{F}^{2} & \lesssim (r(\X)+dJ_y) r, \\
 \int_{\mathcal{T}}\int_{\mathcal{S}} \left\| \left( \widehat \C \left( s,t \right) -\C_0 \left( s,t\right)  \right)\x(s) \right\|^2 dsdt & \lesssim (r(\X)+dJ_y) r,
\end{align*}
where $\theta > 0$ is a positive constant. Here $\lesssim$ means that the inequality holds up to some multiplicative numerical constants.
\end{theorem}
%The proof of Proposition \ref{THM1} is provided in Appendix. %Theorem \ref{THM1} tells us that the error bound of NRRR is at least comparable to that of RRR. 
Theorem \ref{THM1} shows that the prediction error bounds of NRRR are at least comparable to those of reduced-rank regression \citep{bunea2011}. The proof of Theorem \ref{THM1} is in Appendix A. This result provides support for using NRRR in problems with diverging dimensionality; indeed, we see from numerical studies that NRRR always outperforms RRR. We expect that the optimal rate for NRRR is faster than that is given above, since the number of free parameters in a nested low-rank structure can be much smaller than that in a regular reduced-rank structure due to the global dimension reduction by $(\V_0,\U_0)$; see the formulation of the degrees of freedom in \eqref{eq:df} and the discussion afterwards. We will explore this conjecture in our future work.

%From the following simulation part, we can find some facts to support the conclusion here. Under the model settings which contain both global and local level low-dimensional structures, the method with $\U \neq \I_d$ and the one with $\U=\I_d$ have little difference in estimation of $\C$. But adding a dimension reduction regularization on a set of responses can facilitate the interpretation of the model.

\section{Simulation}\label{simu}
%\subsection{Settings and Evaluation Metrics}

We compare the performance of the proposed nested reduced-rank regression (NRRR) methods with several competing methods, including the ordinary least squares method (OLS), the classical reduced-rank regression (RRR), and the reduced-rank ridge regression (RRS). For NRRR, beside the regular version, we consider the special case of setting $r_y=d$, denoted as NRRR-X, and the nested reduced-rank ridge regression, denoted as NRRS, in which a ridge penalty is added to the NRRR criterion for inducing parameter shrinkage.

To generate synthetic data, we let $\x(s) = \{\I_p\otimes \bPhi\trans(s)\}\x$ and 
%\y(t) & = \{\I_d\otimes \bPsi\trans(t)\}\y,\\
$\beps(t) = \{\I_d\otimes \bPsi\trans(t)\}\beps$, where $\x\in \mathbb{R}^{J_xp}$, $\y\in \mathbb{R}^{J_yd}$, and $\beps\in \mathbb{R}^{J_yd}$ are random vectors, and $\bPhi(s)$ and $\bPsi(t)$ are the same two sets of B-spline basis functions used to expand $\C(s,t)$. 
The $\y(t)$ is then given according to \eqref{eq:model2}, i.e., 
$\y(t) = \left\{\I_d\otimes \bPsi\trans(t)\right\}\left\{(\U_0\otimes\I_{J_y})\A_0^*\B_0^{*\trans}(\V_0\otimes\I_{J_x})\trans(\I_{p}\otimes\J_{\phi\phi})\x + \beps\right\}$. Then, for each $i=1,...,n$, the discrete-time observations $(\x_{i}(s), \y_{i}(t))$ are generated as follows,
\begin{enumerate}
    \item Generate $\x_{i}(s)=\left\{\I_p\otimes\bPhi\trans(s)\right\}\x_{i}$ for uniformly distributed time points $s_{u}$, $u=1,\ldots,g$ in $\mathcal{S} = [0,1]$, where $\x_i\in \mathbb{R}^{J_xp}$ is generated from $N(\boldsymbol{0},\bSig)$, where $\bSig=(\rho^{|i-j|})$ with some $0<\rho<1$.
    \item Generate the entries of $\beps_i\in\mathbb{R}^{J_yd}$ as independent samples from $N(0,\sigma^2)$.
    \item Generate
    $\y_{i}(t) = \left\{\I_d\otimes \bPsi\trans(t)\right\}\left\{(\U_0\otimes\I_{J_y})\A_0^*\B_0^{*\trans}(\V_0\otimes\I_{J_x})\trans(\I_{p}\otimes\J_{\phi\phi})\x_i + \beps\right\}$ for uniformly distributed time points $t_{v}$, $v=1,\ldots,m$ in $\mathcal{T} = [0,1]$.
\end{enumerate}
(Here for simplicity, data on different subjects are generated on the same sets of time points.)
%We set $g_i=m_i=100$ for $i=1,\ldots,n$ in all simulation examples. 
The entries of $\A_0^*\in\mathbb{R}^{J_yr_y\times r}$ and $\B_0^*\in\mathbb{R}^{J_xr_x\times r}$ are independent samples from $N(0,1)$, and $\U_0\in\mathbb{R}^{d\times r_y}$ and $\V_0\in\mathbb{R}^{p\times r_x}$ are generated by orthogonalizing random matrices of independent $N(0,1)$ entries via QR decomposition. 

Two settings of model dimensions are considered:
\begin{itemize}
    \item[Setting 1]: $n=100$, $m=g=60$, $p=10$, $d=10$, $r=5$, $r_x=3$, $j_x=8$, $r_y=3$, $j_y=8$.
    \item[Setting 2]: $n=100$, $m=g=100$, $p=20$, $d=20$, $r=3$, $r_x=3$, $j_x=8$, $r_y=3$, $j_y=8$.
\end{itemize}
In Setting 1, the model dimensions, $p j_x = 80,\ dj_y = 80$ are comparable and a bit smaller than the sample size; but the number of unknowns, $80\times 80$, is already very large. In Setting 2, the model dimensions are much higher than the sample size, i.e., $p j_x = 160,\ dj_y = 160$, and the total number of unknowns is four times of that in Setting 1. 
For each setting, we try different signal to noise ratios ($\mbox{SNR}\in \{1,2,4\}$), defined as the ratio between the standard deviation of all the elements in the response matrix
$
(\U_0\otimes\I_{J_y})\A_0^*\B_0^{*\trans}(\V_0\otimes\I_{J_x})\trans(\I_{p}\otimes\J_{\phi\phi})(\x_1,\x_2,\ldots,\x_n)$ and the noise level $\sigma$, and different design correlations ($\rho \in \{0.1,0.5,0.9\}$). The ranks and other tuning parameters (if there is any) are selected by 10-fold cross validation. For methods with nested reduced-rank structure, we use the proposed BIC criterion to select ranks. The experiment is replicated 300 times for each setting. %(The results displayed are the trimmed version with the smallest and largest 20 observations deleted).

% \eqref{eq:bic}. %A summary of all the error measurements and the estimated ranks $r,r_x,r_y$ is reported. Some of the results are in the supplementary part.

To evaluate the performance of different methods, we compute for each method % the mean squared estimation error (MSEE) as 
% $$
% \text{MSEE}(\widehat \C, \C)=\frac{1}{p d J_x J_y}\|\widehat \C-\C\|_{\rm F}^2.
% $$
% and 
the trimmed mean squared prediction error (MSPE) from all runs (the smallest and largest 20 observations are deleted from 300 runs), 
\begin{align*}
\text{MSPE}(\widehat \C, \C_0)=\frac{1}{n_{te}}\|\Y_{te}-\X_{te}\widehat \C\|_{\rm F}^2%\label{eq:MSPE}
\end{align*}
based on independent testing set of size $n_{te} = 500$, where $\Y_{te}$ and $\X_{te}$ are the integrated response and predictor matrices. Similarly, to evaluate the estimation of the functional responses, we compute the trimmed mean squared functional prediction error (MSFPE),  
%{\color{blue} 
\begin{align*}
% \text{MSFPE}(\widehat \y,\y)=\frac{1}{n_{te}}\sum_{i=1}^{n_{te}} \int \|\y_{te,i}(t)-\widehat \y_{te,i}(t)\|_{\rm F}^2dt.\label{eq:MSFPE}
\text{MSFPE}(\widehat \y,\y)=\frac{1}{n_{te}}\sum_{i=1}^{n_{te}} \sum_{v=1}^{m} \|\y_{te,i}(t_v)-\widehat \y_{te,i}(t_v)\|_{\rm F}^2.%\label{eq:MSFPE}
\end{align*}
%All the integrals are computed as the Riemann sums with the discrete observations uniformly distributed over $\mathcal{S} = \mathcal{T} = [0,1]$.
%}

%\subsection{Results}

Table \ref{tab:ypred1} and Table \ref{tab:ypred2} present the prediction errors (MSPE) under Settings 1 and 2, respectively. The results from OLS are omitted as they are much worse than those of the other methods. Among the five methods presented, RRR has the worst performance. The performance of NRRR is slightly better than that of NRRR-X. RRS substantially improves its corresponding counterpart RRR by incorporating $\ell_2$ shrinkage estimation. In general the improvement is more substantial when the SNR is low and/or the design correlation is high. In contrast, in most scenarios NRRS only slightly outperforms or just has comparable performance to NRRR. This is because NRRR has already considered a finer low-dimensional structure so that the extra shrinkage becomes less effective. Due to space limit, we present the results on estimating $r$, $r_x$ and $r_y$ in Appendix B. RRR usually leads to underestimation of $r$; this is expected as RRR tries to use an overall low-rank structure to mimic the finer or even lower dimensional nested low-rank structure. NRRR methods perform well in rank estimation in general. Therefore, the results confirm that NRRR can produce a more interpretable model with improved predictive accuracy.

%outperforms NRRR a little bit when random noise is large, but when random noise becomes smaller they both perform well. Under setting 2, NRRR and NRRS are comparable to each other.

\begin{table}[htbp]
  \centering
  \caption{Simulation results for Setting 1. The mean MSPE values are reported with their standard deviations in parentheses. To improve presentation, all values are multiplied by 10.}\label{tab:ypred1}
  \begin{tabular}{lrrrrrr}
    \hline
      & \multicolumn{1}{c}{$\rho$} & \multicolumn{1}{c}{NRRR} & \multicolumn{1}{c}{NRRR-X} & \multicolumn{1}{c}{RRR} & \multicolumn{1}{c}{RRS} & \multicolumn{1}{c}{NRRS}\\\hline
    \hfill      & 0.1 & 11.43 (2.64) & 12.16 (2.81) & 14.47 (3.16) & 11.34 (2.46) & 10.97 (2.50) \\ 
  {$\mbox{SNR}=1$} \ & 0.5 & 18.14 (4.28) & 19.07 (4.33) & 22.42 (4.92) & 17.46 (3.81) & 17.61 (4.18) \\ 
  \hfill        & 0.9 & 26.20 (9.17) & 26.58 (9.01) & 29.56 (9.92) & 23.87 (8.04) & 25.6 (8.92) \\ \hline
  \hfill        & 0.1 & 2.68 (0.56) & 2.84 (0.59) & 3.84 (0.80) & 3.08 (0.59) & 2.77 (0.55) \\ 
  {$\mbox{SNR}=2$} \ & 0.5 & 4.18 (1.01) & 4.47 (1.10) & 5.91 (1.40) & 4.56 (1.06) & 4.20 (0.99) \\ 
  \hfill        & 0.9 & 6.42 (2.19) & 6.79 (2.31) & 8.26 (2.58) & 6.48 (2.09) & 6.29 (2.06) \\ \hline
  \hfill        & 0.1 & 0.65 (0.14) & 0.68 (0.15) & 0.92 (0.21) & 0.96 (0.19) & 0.77 (0.17) \\ 
  {$\mbox{SNR}=4$} \ & 0.5 & 1.04 (0.26) & 1.08 (0.27) & 1.47 (0.38) & 1.31 (0.29) & 1.14 (0.27) \\ 
  \hfill        & 0.9 & 1.52 (0.52) & 1.61 (0.55) & 2.11 (0.70) & 1.69 (0.53) & 1.59 (0.51) \\ 
   \hline
  \end{tabular}
\end{table}

\begin{table}[htbp]
  \centering
  \caption{Simulation results for Setting 2. The layout is the same as in Table \ref{tab:ypred1}. 
  }\label{tab:ypred2}
  \begin{tabular}{lrrrrrr}
    \hline
     & \multicolumn{1}{c}{$\rho$} & \multicolumn{1}{c}{NRRR} & \multicolumn{1}{c}{NRRR-X} & \multicolumn{1}{c}{RRR} & \multicolumn{1}{c}{RRS} & \multicolumn{1}{c}{NRRS}\\\hline
   \hfill       & 0.1 & 6.20 (1.47) & 6.56 (1.55) & 7.82 (1.98) & 6.97 (1.60) & 6.30 (1.50) \\ 
  {$\mbox{SNR}=1$} \ & 0.5 & 9.76 (3.15) & 10.33 (3.38) & 11.82 (3.91) & 10.55 (3.29) & 9.76 (3.12) \\ 
  \hfill        & 0.9 & 14.44 (5.72) & 15.06 (5.87) & 16.21 (6.40) & 14.88 (5.76) & 14.28 (5.67) \\ \hline
  \hfill        & 0.1 & 1.56 (0.41) & 1.58 (0.41) & 2.40 (1.11) & 2.07 (0.49) & 1.61 (0.43) \\ 
  {$\mbox{SNR}=2$} \ & 0.5 & 2.46 (0.74) & 2.51 (0.74) & 3.16 (0.98) & 3.08 (0.88) & 2.49 (0.73) \\ 
  \hfill        & 0.9 & 3.28 (1.22) & 3.40 (1.28) & 3.86 (1.46) & 3.91 (1.66) & 3.34 (1.22) \\  \hline
  \hfill        & 0.1 & 0.37 (0.10) & 0.38 (0.10) & 1.05 (1.01) & 0.78 (0.18) & 0.42 (0.16) \\ 
  {$\mbox{SNR}=4$} \ & 0.5 & 0.61 (0.19) & 0.62 (0.19) & 0.95 (0.28) & 0.91 (0.23) & 0.63 (0.19) \\ 
  \hfill        & 0.9 & 0.88 (0.35) & 0.90 (0.36) & 1.05 (0.41) & 1.07 (0.44) & 0.89 (0.37) \\ 
   \hline
  \end{tabular}
\end{table}

\begin{figure}[htp]
\centering 
\subfigure[Setting 1]{\includegraphics[width=0.48\textwidth]{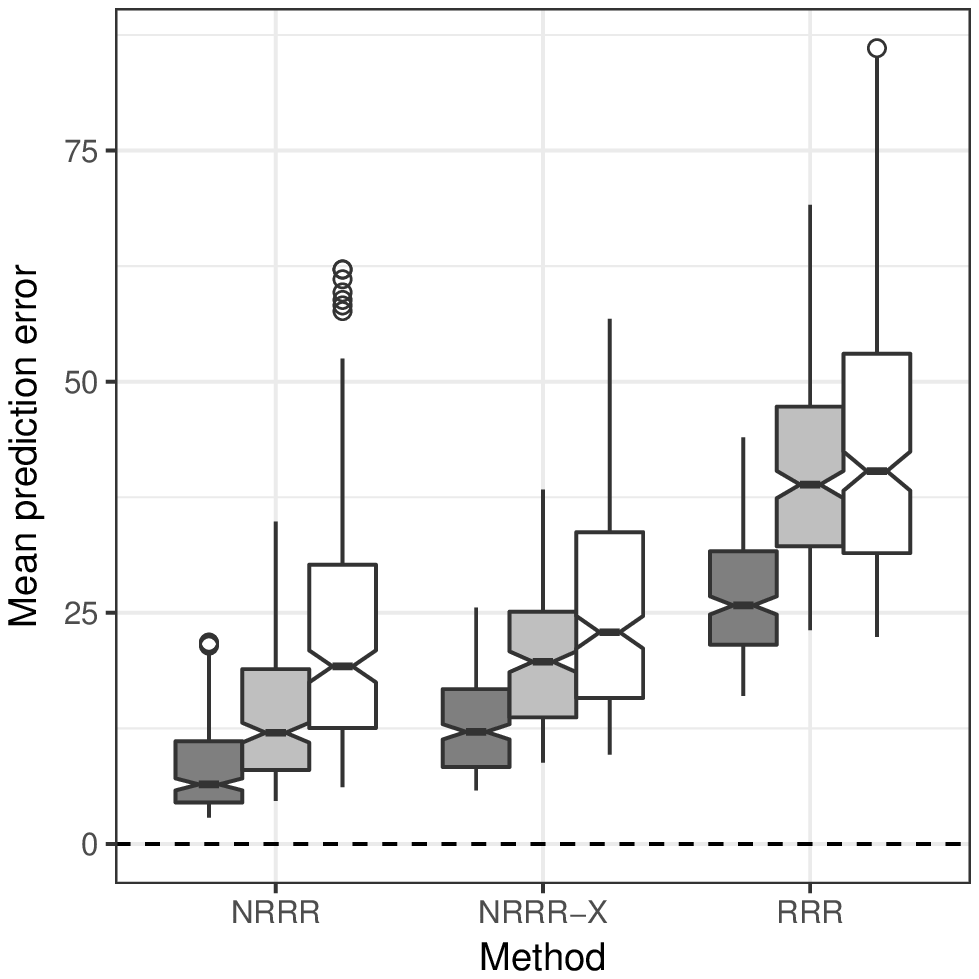}}
\subfigure[Setting 2]{\includegraphics[width=0.48\textwidth]{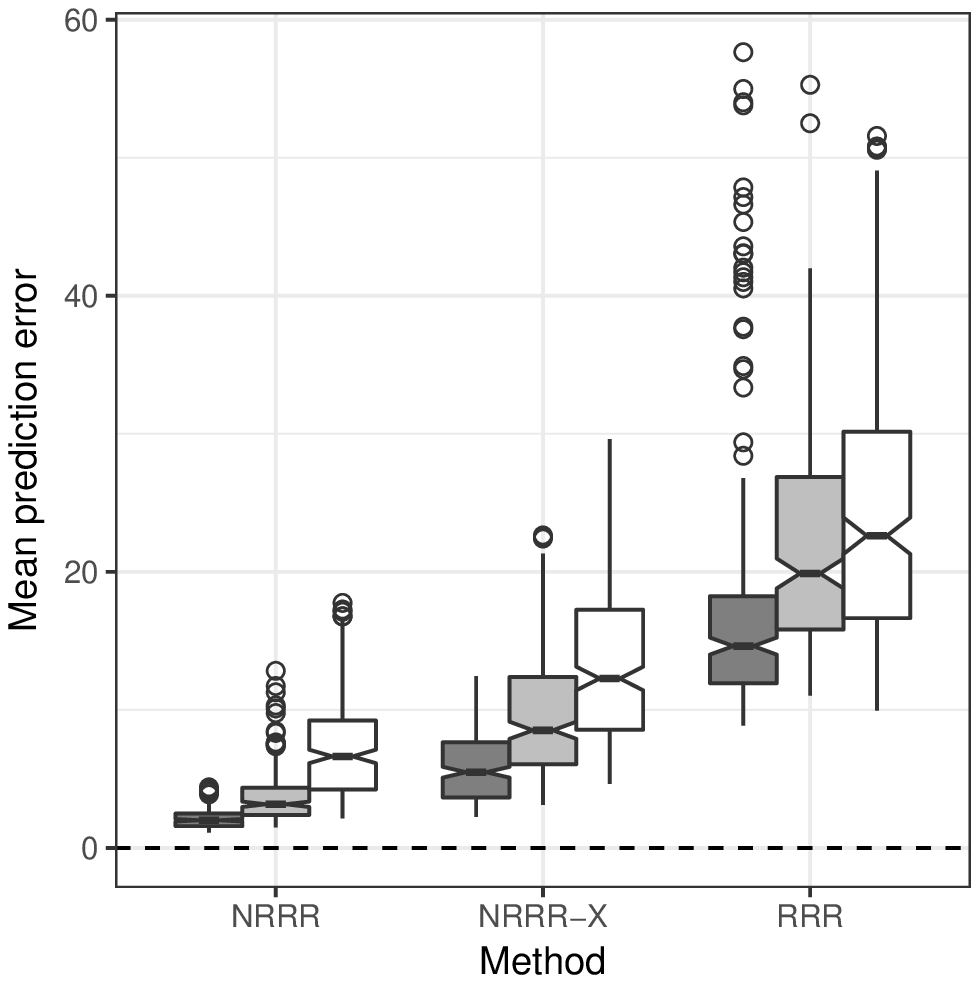}}
\caption{Boxplots of MSFPE from 300 simulation runs. In each panel, each set of three boxplots for $\rho = 0.1, 0.5, 0.9$ is showing in black, grey and white colors from left to right.}\label{fig:boxplot1}
\end{figure}

\begin{figure}[htp]
\includegraphics[width=.48\textwidth]{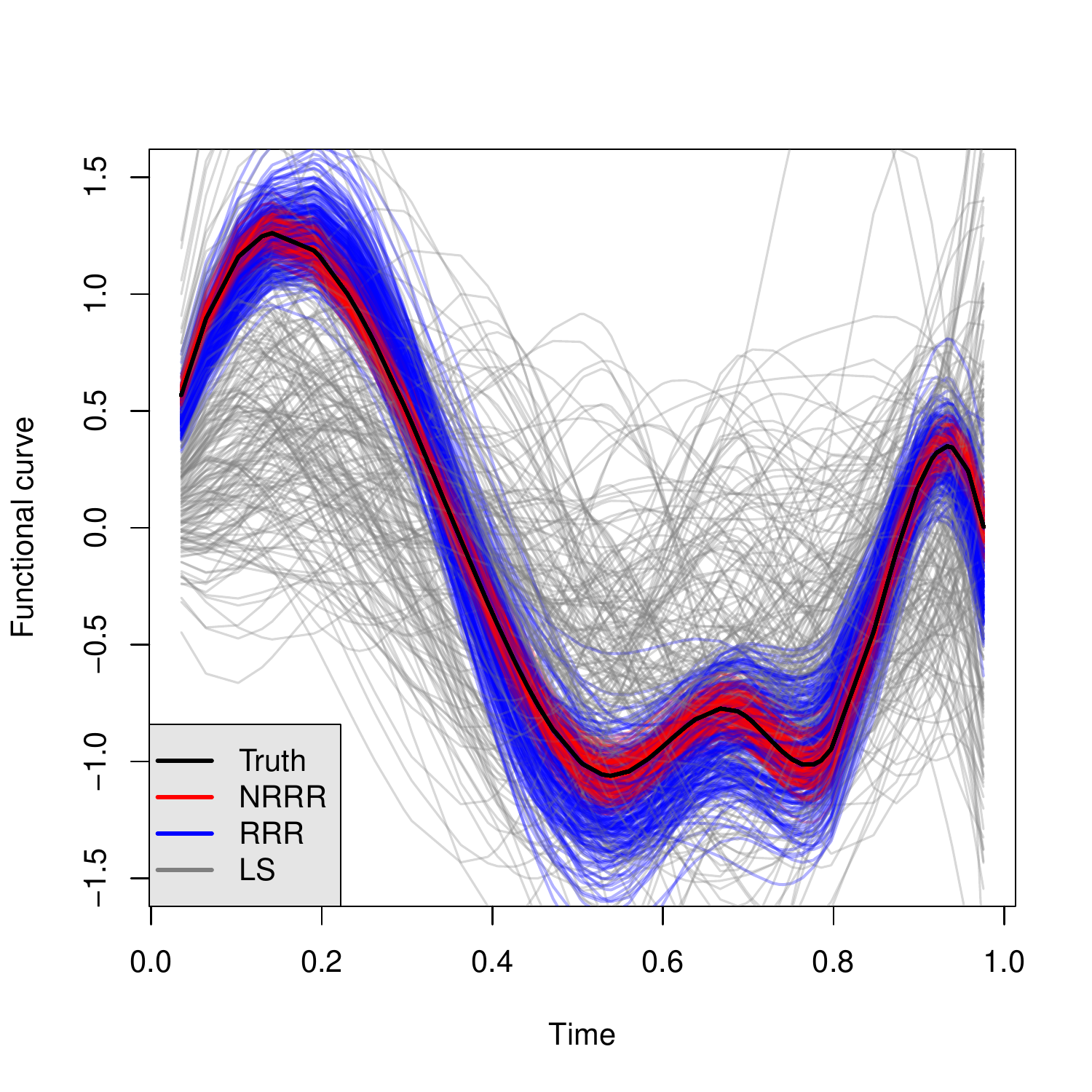}\hfill
\includegraphics[width=.48\textwidth]{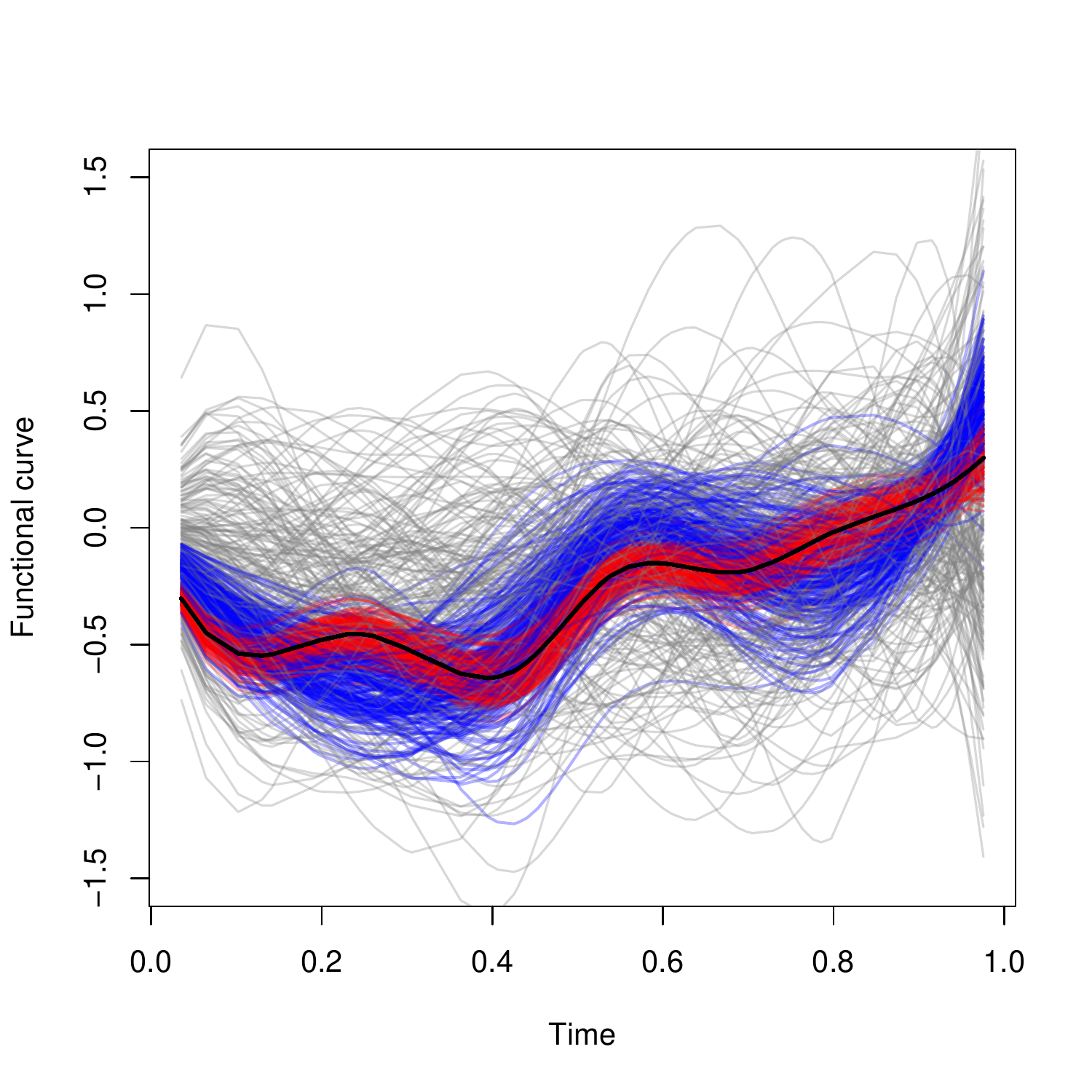}\hfill
\caption{Comparison of the true curves and the predicted curves in two simulation runs under Setting 1 with $\mbox{SNR} =2$ and $\rho=0.5$.}\label{fig:curve}
\end{figure}

To visualize the effects of nested low-rank dimension reduction, Figure \ref{fig:boxplot1} displays the boxplots of MSFPE for NRRR, NRRR-X and RRR under Settings 1 and 2 with $\mbox{SNR} = 1$, and Figure \ref{fig:curve} draws two particular sets of the true and predicted curves by NRRR, RRR and OLS from the simulation. The efficacy of the nested dimension reduction is apparent. The results under other settings deliver the same message and hence are omitted. Except for RRR and RRS, all the above results are obtained from using BIC to select the model ranks. The results obtained from using 10-fold cross validation for all methods are similar and presented in Appendix B.

\section{Application to Adelaide Electricity Demand Data}\label{AppAde}

Adelaide is the capital city of the state of South Australia. %and has more than 77\% of its population. 
The city has a Mediterranean climate, with warm-dry summers and cool-mild winters. In summer time, the cooling mainly depends on air conditioning, which makes the electricity demand highly dependent on the weather conditions, and a large volatility in temperature throughout the day could make stable electricity supply challenging. Therefore, it is of great interest to understand the dependence and the predictive association between the electricity demand and the temperature, for facilitating the supply management of electricity \citep{magnano2007,magnano2008generation,shufan2015}. Here we apply NRRR to perform a multivariate functional regression analysis between daily half-hour electricity demand profiles for the 7 days of a week and the corresponding temperature profiles for the 7 days of the same week.

\begin{figure}[htp]
\subfigure[]{\includegraphics[width=.32\textwidth]{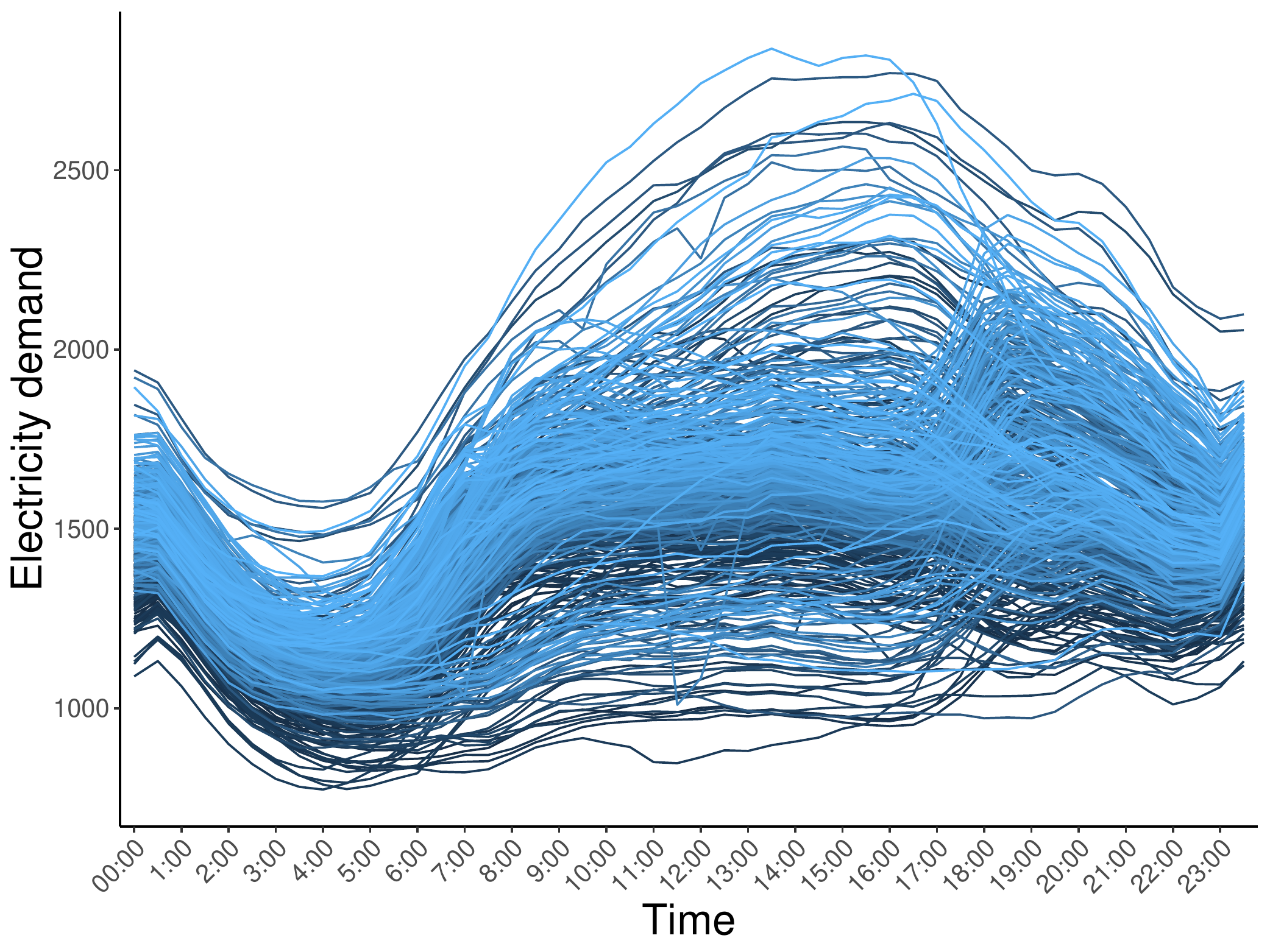}}
\subfigure[]{\includegraphics[width=.32\textwidth]{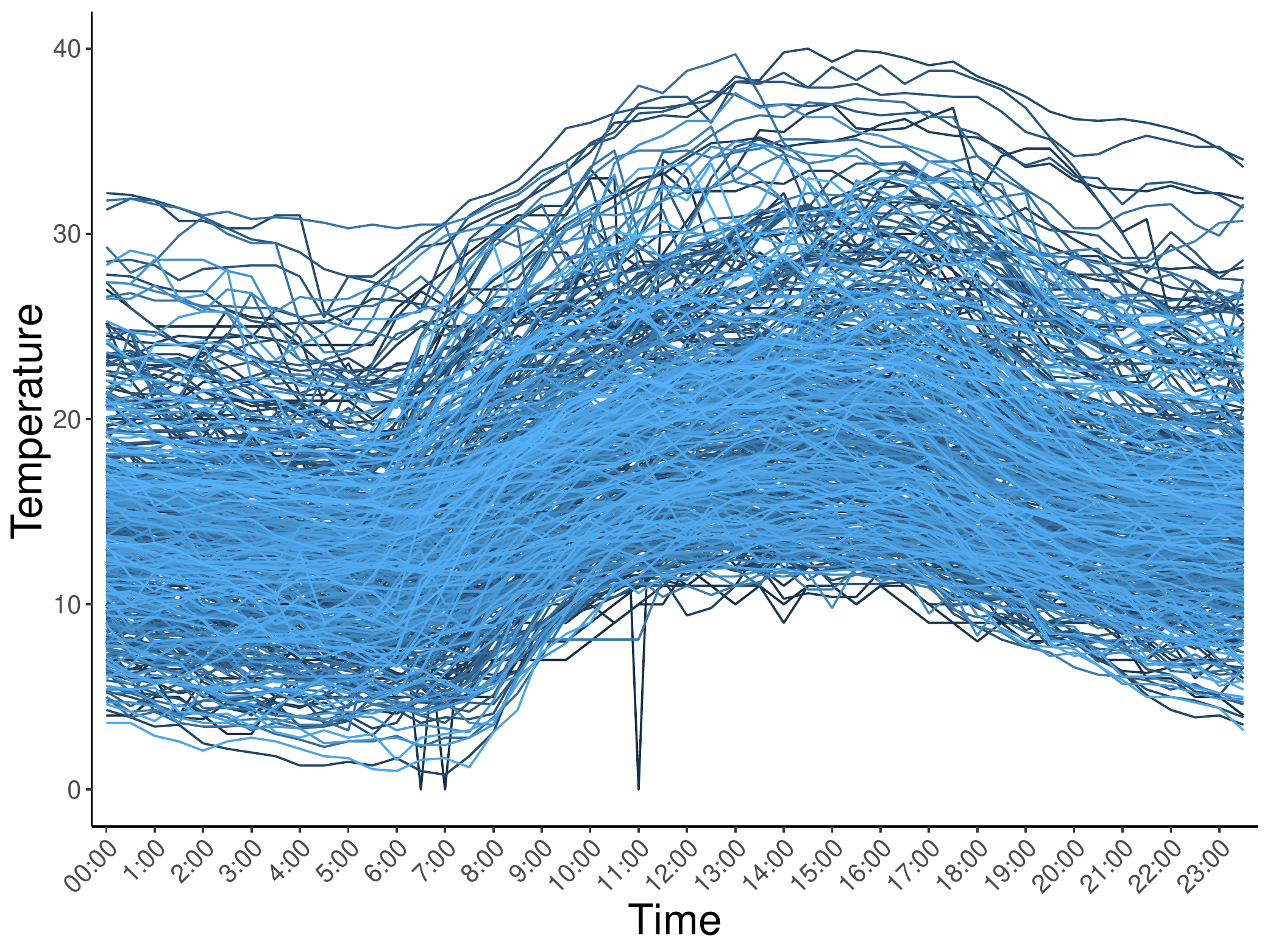}}
\subfigure[]{\includegraphics[width=.32\textwidth]{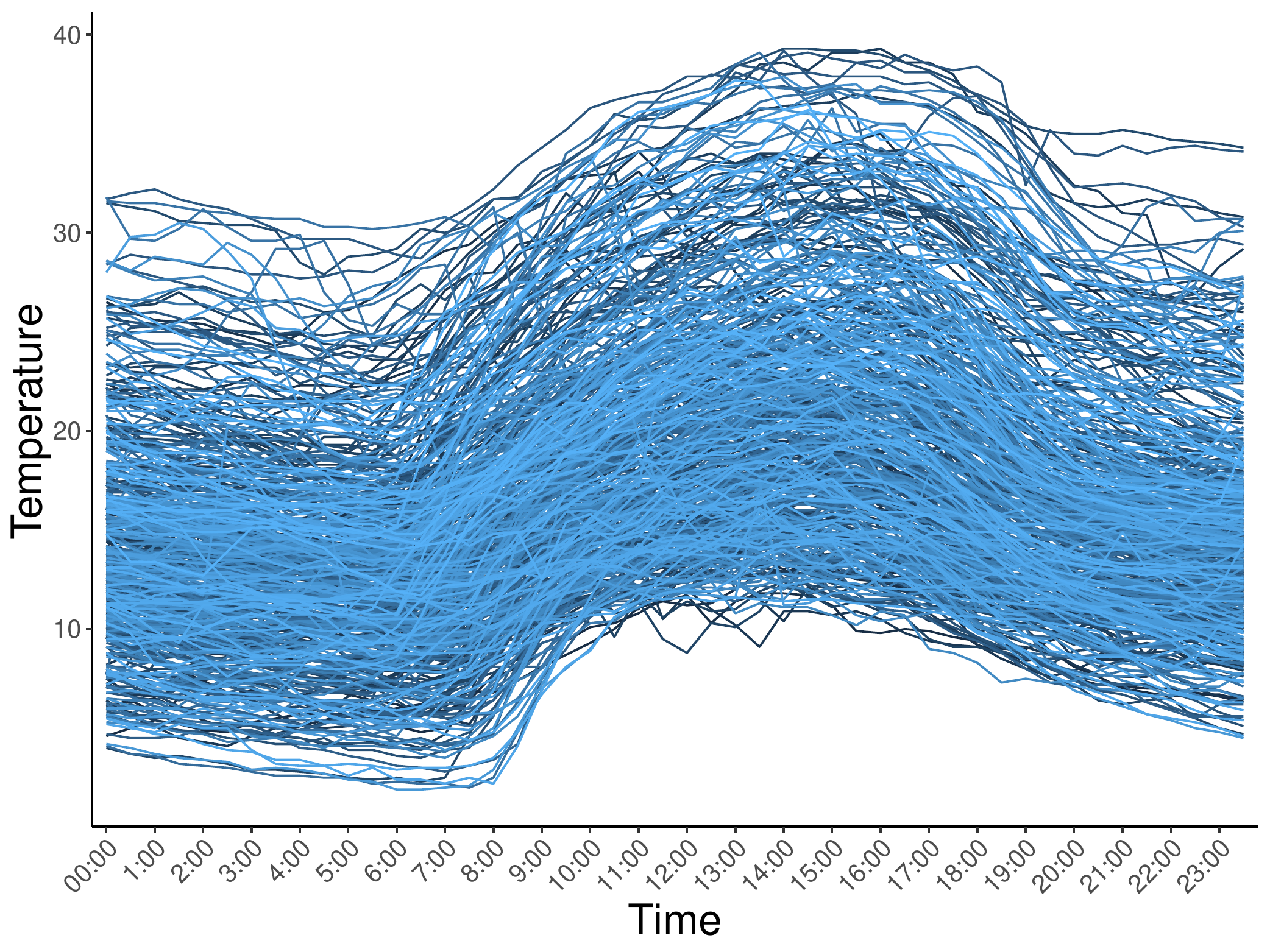}}
\caption{Adelaide electricity demand analysis: (a) electricity demand in Adelaide, (b) temperature in Kent town, and (c) temperature in the airport. Plotted are the half-hourly observed profiles for all Mondays.}\label{Ade:Monday}
\end{figure}

Half-hourly temperature records at two locations, Adelaide Kent town and Adelaide airport, are available between 7/6/1997 and 3/31/2007. Also available are the half-hourly electricity demand records of Adelaide for the same period. As such, for each day during the period, there are three observed functional curves, each with 48 half-hourly observations. As an illustration, Figure \ref{Ade:Monday} plots the temperature and electricity demand profiles of all the Mondays from 7/6/1997 to 3/31/2007. Since our main focus is on studying the general association between the within-day demand and temperature trajectories in a week, we center the 48 discrete observations of each daily curve, to remove the between-day trend and seasonality of the data. Each week is then treated as a replication. %We remark that alternatively more sophisticated time series models could be used for the data pre-processing.  

After data pre-processing, we use the daily half-hour electricity demand as the functional multivariate response with $d=7$ (corresponding to 7 days in a week from Monday to Sunday), and as for the predictors, we consider two settings. In the first setting, we only use the half-hour temperature data from Kent as the multivariate functional predictors, so that $p=7$; in the second setting, we also include temperature data from the airport to make $p=14$. Not surprisingly, the two sets of temperature data are extremely highly correlated, so the second setting is meant to test for the behaviors of different methods in the presence of high collinearity. In either setting, the total sample size is $n=508$, equaling to the number of weeks in the study period. To leave sufficient flexibility in estimating the regression surface, we use B-spline with 30 degrees of freedom to convert the discrete observations to its integrated form according to \eqref{eq:y}.

\begin{figure}
\includegraphics[width=.31\textwidth]{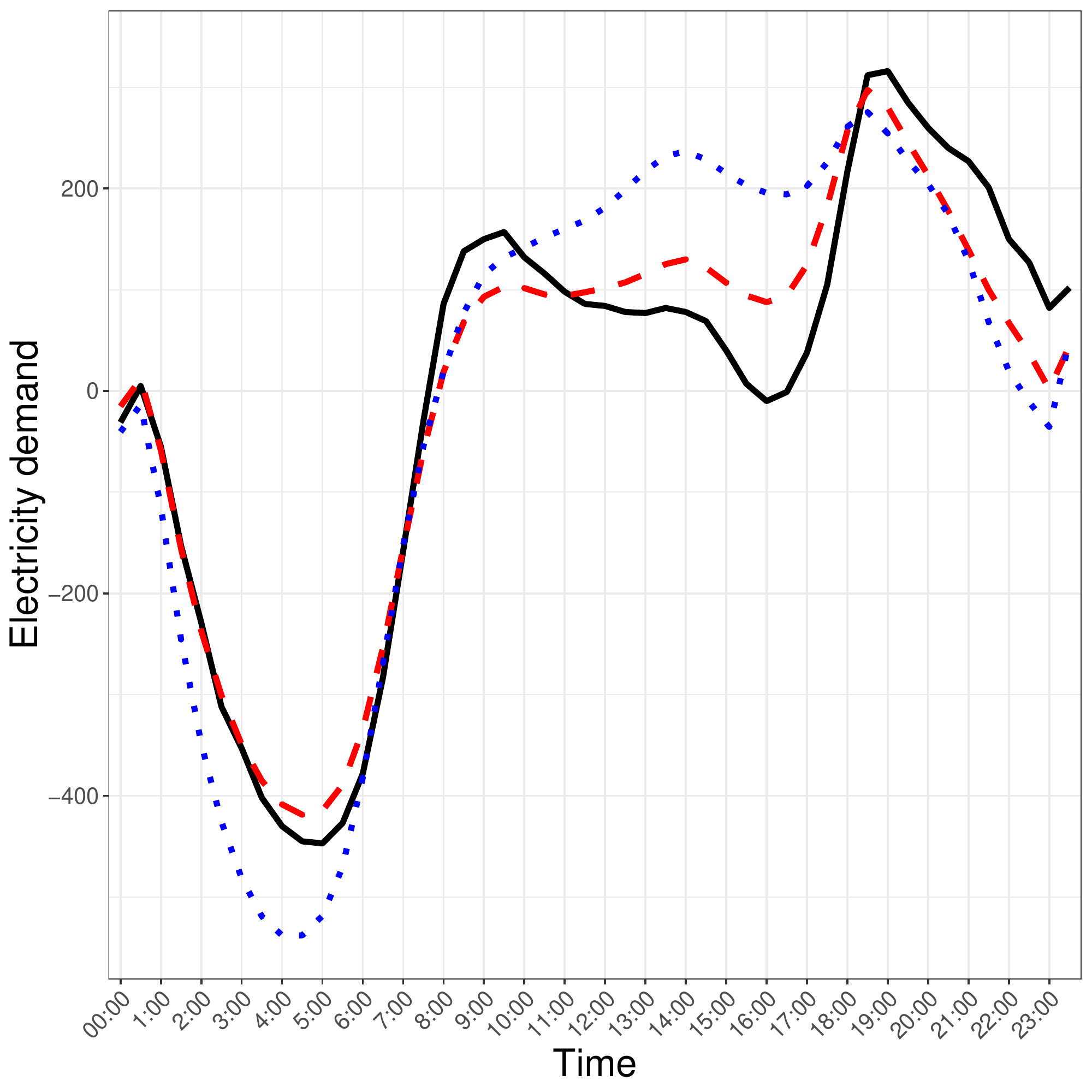}\hfill
\includegraphics[width=.31\textwidth]{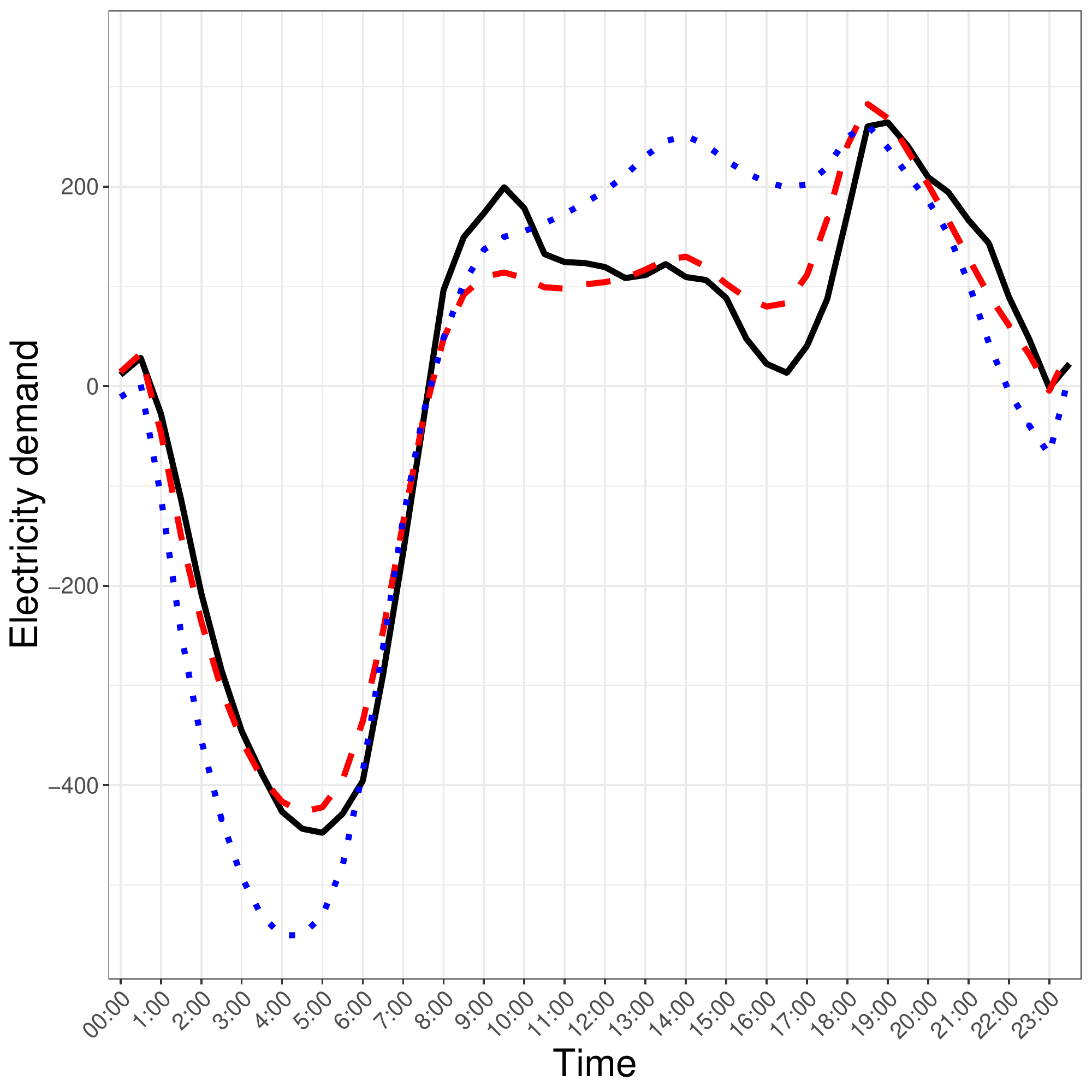}\hfill
\includegraphics[width=.31\textwidth]{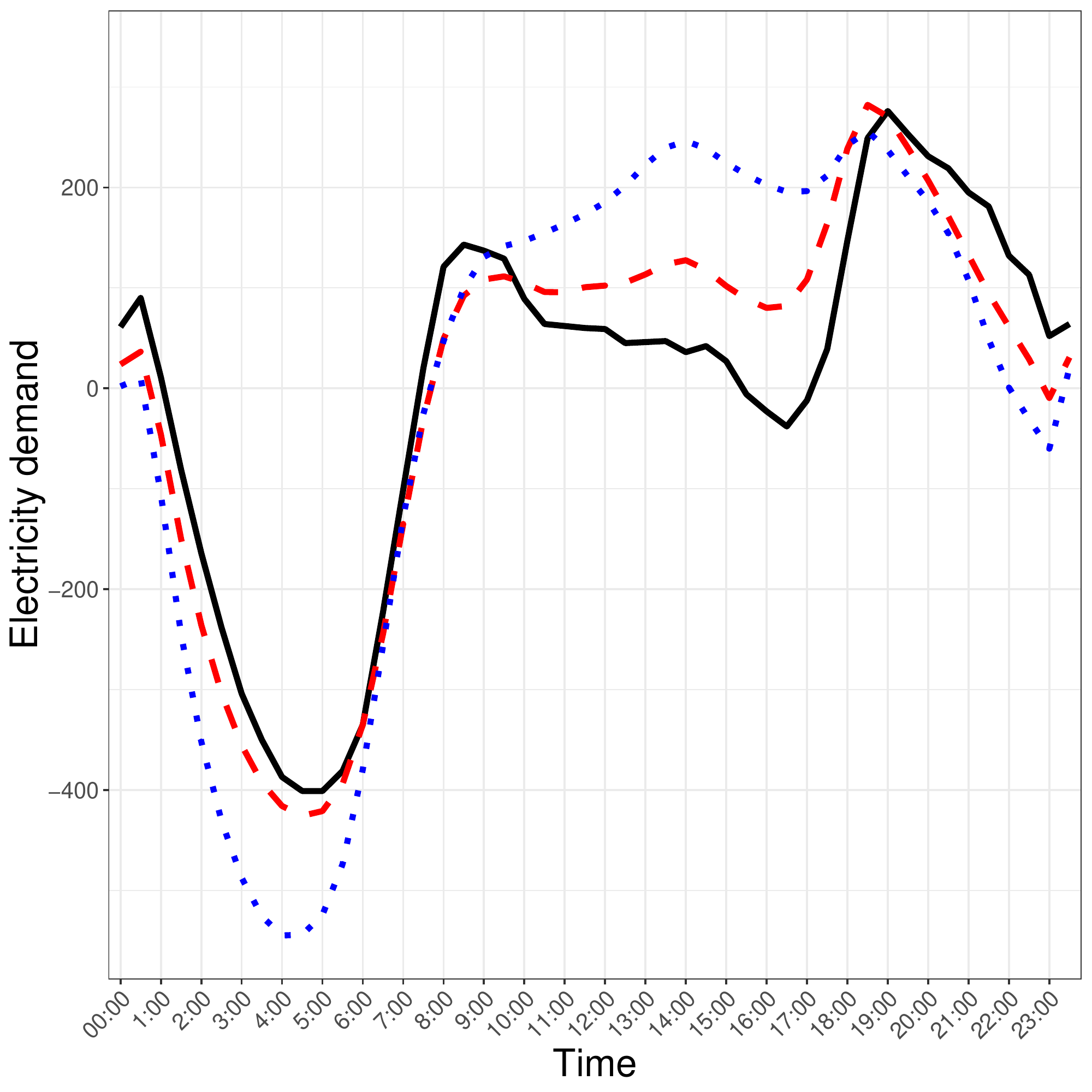}\hfill
\caption{Adelaide electricity demand analysis: randomly selected examples of observed demand curves (in black) and out-of-sample predicted curves by RRR (in blue) and NRRR (in red).}\label{fig:randomsplit}
\end{figure}

First, we compare different methods using an out-of-sample random splitting procedure. Each time, we randomly select 400 samples as the training set and the remaining 108 samples as the test set. The model is fitted using the training data, and the relative mean squared prediction error (RMSPE) is then computed based on the test data,
%{\color{blue}
$$
\mbox{RMSPE}(\widehat \y,\y)=\frac{1}{n_{te}}\sum_{i=1}^{n_{te}} \frac{\int \|\y_{te,i}(t)-\widehat\y_{te,i}(t)\|^2dt}{\int \|\y_{te,i}(t)\|^2dt}.
$$
%}
The procedure is repeated 100 times, and the results are reported in Table \ref{tab:randomsplit}. In both settings, NRRR and NRRS perform very well, and their predicted curves are able to account for about 74\% of the total variation in the observed demand curves. The results show that there is a dramatic global dimension reduction of the functional predictors, as $r_x$ is estimated to be only $1$ most of the times. As $r_y$ is often close to the number of original functional responses, this indicates that each daily electricity demand curve  has its own pattern, and thus there is not much room for a global dimension reduction. In contrast, RRR and RRS perform much worse in prediction, and RRR even fails completely in Setting 2. To visualize, Figure \ref{fig:randomsplit} plots some randomly selected observed and predicted curves under Setting 1; the superior performance of NRRR is apparent. These results clearly show the power and necessity of global dimension reduction, especially in the presence of high correlation among the functional predictors.  

%The predicted curves can account for a large portion of variation in the observed curves.

%{\color{red} Xiaokang, I am confused about the results in this table. What is this reporting? Under setting 1 or setting 2? How is this different from the results reported in the following subsections?}

\begin{table}[ht]
\label{tab:randomsplit}
\centering
\caption{Adelaide electricity demand analysis: out-of-sample performance of different methods. Reported are the means and standard deviations (in parenthesis) of RMSPE, $r$, $r_x$ and $r_y$ over 100 simulation runs.} 
\begin{tabular}{rlllll}
  \hline
 & Methods & RRR & NRRR & RRS & NRRS \\ 
  \hline
Setting 1 & RMSPE & 0.42 (0.04) & 0.27 (0.02) & 0.38 (0.03) & 0.26 (0.02) \\ 
  \hfill & $r$ & 1.53 (0.63) & 4.06 (0.28) & 3.60 (0.70) & 4.09 (0.35) \\ 
  \hfill & $r_x$ & \hfill & 1.00 (0.00) & \hfill & 1.01 (0.10) \\ 
  \hfill & $r_y$ & \hfill & 5.70 (1.47) & \hfill & 5.73 (1.43) \\ 
  Setting 2 & RMSPE & 1.08 (0.19) & 0.26 (0.02) & 0.55 (0.05) & 0.26 (0.02) \\ 
  \hfill & $r$ & 0.26 (0.44) & 4.45 (0.89) & 1.00 (0.00) & 4.40 (0.80) \\ 
  \hfill & $r_x$ & \hfill & 1.00 (0.00) & \hfill & 1.01 (0.10) \\ 
  \hfill & $r_y$ & \hfill & 6.72 (0.57) & \hfill & 6.75 (0.52) \\ 
   \hline
\end{tabular}
\end{table}

% \begin{table}[htbp]
% \centering
% \caption{Relative mean squared prediction error (in $\%$)}\label{tab:SA0}
% \begin{tabular}{l*{2}{c}r}
% Methods & RRR & NRRR \\
% \hline
% mean &22.18 &15.13\\
% sd& 1.12 &1.24\\
% trimmed mean & 22.10 &15.02\\
% trimmed sd &0.49 &0.49\\
% mode &22.12&14.57\\ 
% median &22.07& 14.89\\
% \end{tabular}
% \end{table}

%Table \ref{tab:SA0} reports the average RMSPE for RRR and NRRR. 

%Next, we use the first 500 sample to fit the model and use the remaining 8 samples to illustrate the predictive power of the proposed NRRR method. Here we consider two model settings, the first one only includes the temperature data of Kent town, and the second one regard the temperature data for both Kent town and Adelaide airport as the predictors. For both of two settings, we use the 508th sample to depict the prediction performance. Following are the results for the setting $p=7, d=7$.

%\subsection{Results for Setting 1: Kent Town Temperature Data Included}

We then use all data to fit a final NRRR model with only the temperature observations from Kent. The estimated rank values are  $\widehat{r} = 4$, $\widehat r_x=1$, and $\widehat r_y=5$. The estimated loading matrix for the predictors is $\widehat \V = (0.22, 0.39, 0.46, 0.52, 0.43, 0.28, 0.25)\trans.$
This shows that there is only one latent functional predictor that is driving the patterns of the electronic demands, and this factor can be roughly explained as the averaged daily temperature profile of the week. It appears that the days closer to the middle of the week load higher. On the response side, there is not much global reduction, as the estimated loading matrix $\widehat{\U}$ is of rank 5. To make sense of $\widehat{\U}$, it may be more convenient to examine the two basis vectors of its orthogonal complement, i.e., the first two singular vectors of $\I - \widehat{\U}\widehat{\U}\trans$, which give the latent response factors that are not related to the temperatures at all.
%\[
%\begin{bmatrix}
%-0.52 & 0.00\\
%0.36  & -0.68\\
%0.28  & 0.73\\
%0.25  & -0.00\\
%-0.56 & -0.04\\
%0.34  & 0.05\\
%-0.18 & -0.04
%\end{bmatrix}.
%\]
While the first loading vector $(-0.52,0.36,0.28,0.25,-0.56,0.34,-0.18)\trans$ is hard to interpret, the second loading vector $(0.00,-0.68,0.73,0.00,-0.04,0.05,-0.04)\trans$ clearly indicates that the difference between the electronic demand profiles of Tuesday and Wednesday is mostly a noise process. In other words, the demand profiles of these two days are related to the temperature process in almost the same way.

\begin{figure}
\subfigure[Tuesday]{\includegraphics[width=.45\textwidth]{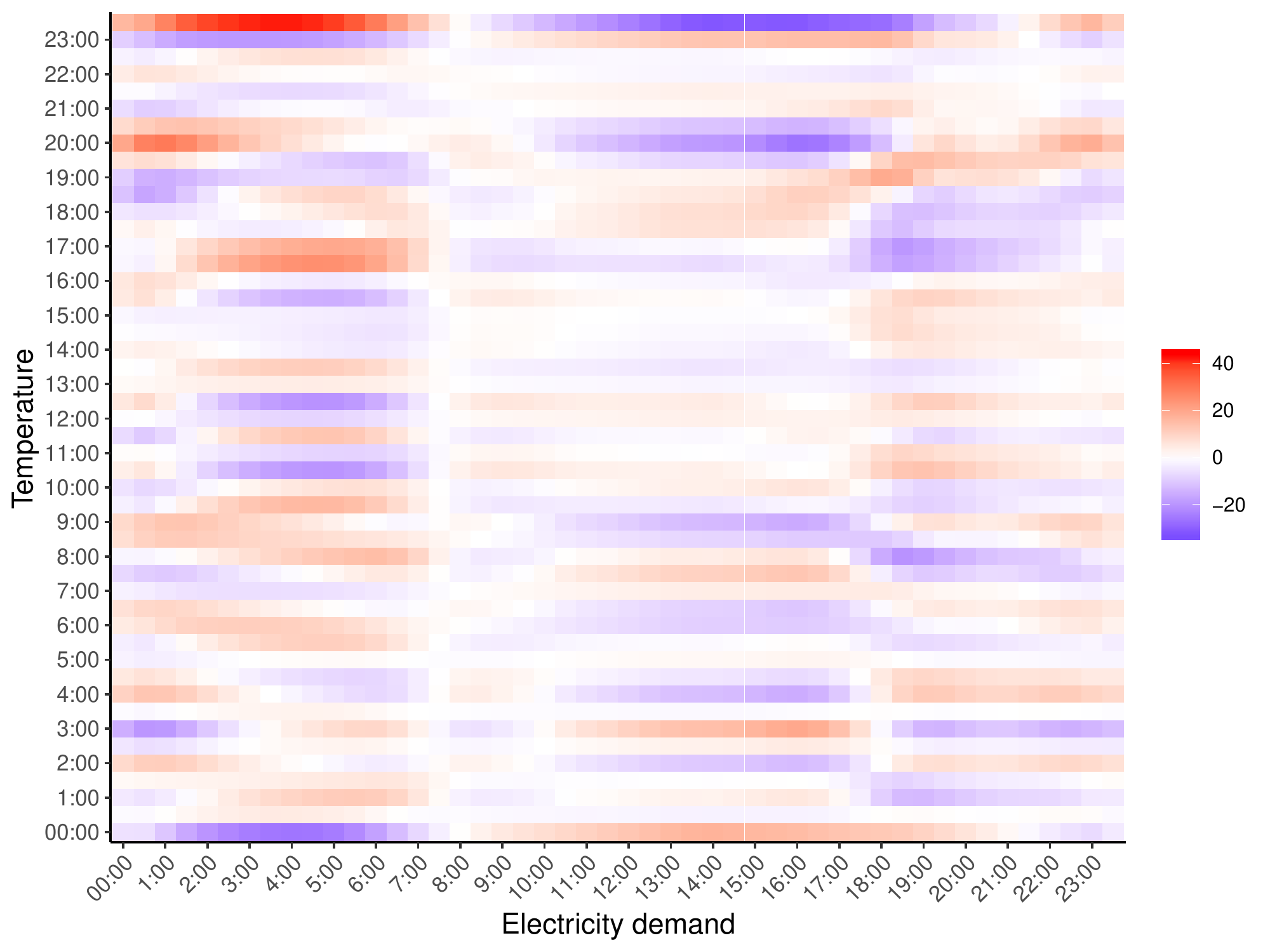}}\hfill
\subfigure[Saturday]{\includegraphics[width=.45\textwidth]{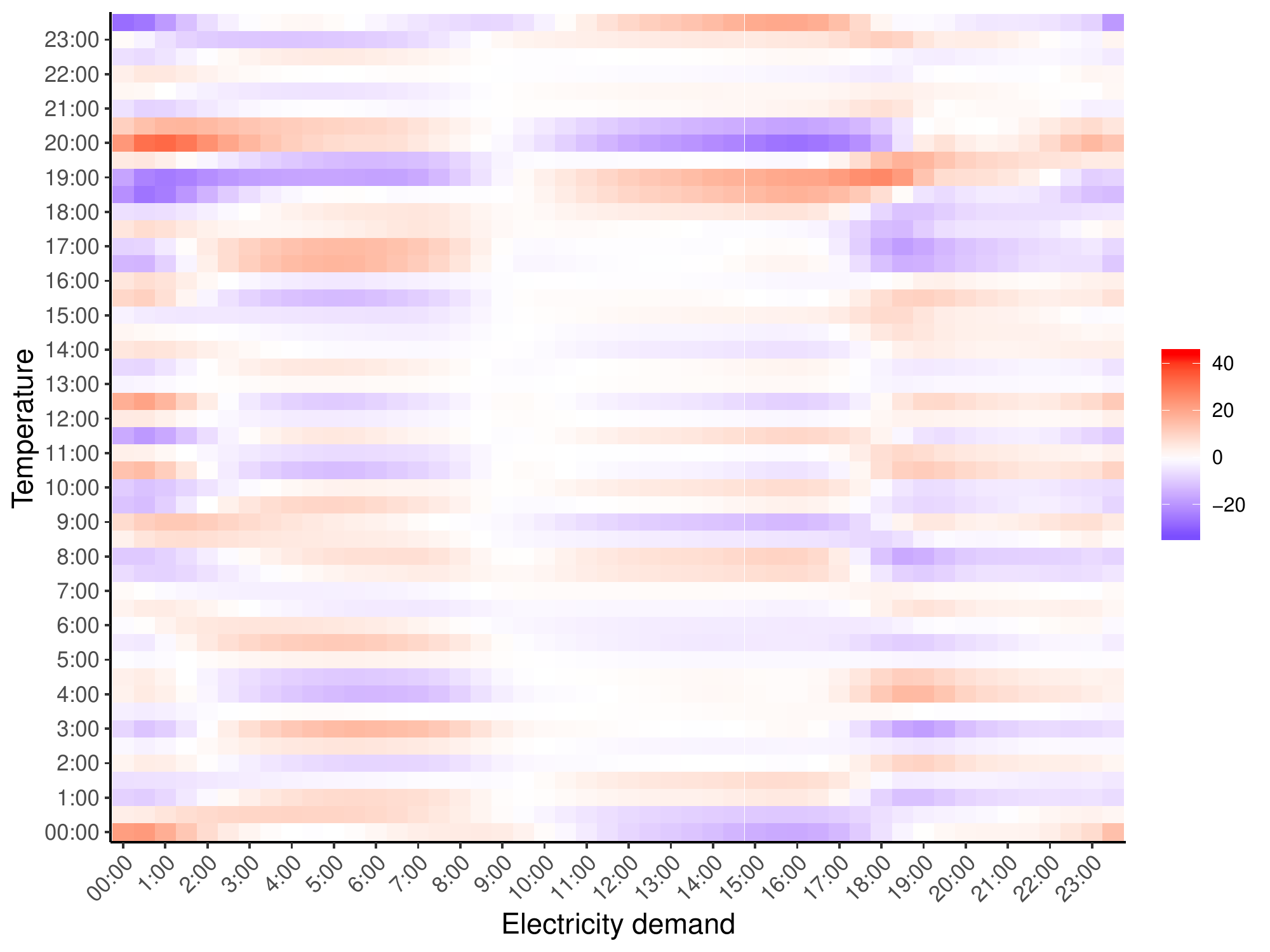}}\hfill
\caption{Adelaide electricity demand analysis: heatmaps of estimated regression surfaces defined in \eqref{eq:heatmapsurface}.}\label{fig:heatmap}
\end{figure}

Let $\widetilde{\u}_k$ be the $k$th row of $\widehat{\U}$. Then Model \eqref{eq:model2} shows that the estimated regression surface
\begin{align}
  \widetilde{c}_k(s,t) = \widetilde{\u}_k\trans(\I_{\widehat{r}_y} \otimes \bPsi\trans(t))\widehat{\A}^* \widehat{\B}^{*\rm T} (\I_{\widehat{r}_x} \otimes \bPhi(s)),\qquad k=1,\ldots,d, 
  \label{eq:heatmapsurface}
\end{align}
would indicate how the response $y_k(t)$ is related to the latent predictor $\widehat{\V}\trans\x(s)$ over $s$ and $t$. In the context of this application, $\widetilde{c}_k(s,t)$ shows that how the electricity demand trajectory on the $k$th day of a week is related to the trajectory of the average temperature of the week. We therefore plot the heatmaps of these surfaces to visualize. Figure \ref{fig:heatmap} displays the plots for Tuesday and Saturday. While the patterns of the association are hard to comprehend in general, some observations can be made. First, there are three association regimes throughout each day, i.e., night hours from about midnight to 7:30, daylight hours from about 7:30 to 18:00, and the rest hours from about 18:00 to midnight. This corresponds well with the general patterns of daily electricity demand, and the three regimes are separated by the ``Morning ramp'', i.e., the transition from relatively lower loads to higher loads in the morning, and the peak load time around 18:00. Noticeably, the electricity demand in daylight hours is the least associated with the temperature. Another observation is that temperatures between about 19:00 to 20:30 and 23:00 to 00:00 in general have the largest effects on the electricity demand. This may be related to household and entertainment activities. Lastly, we observe that the association patterns on the workdays are similar to each other, and are slightly different from those on the weekends. 

\section{Discussion}\label{diss}

There are many research directions that stem from the proposed nested reduced-rank estimation framework. Our method can be extended to the historical functional regression, i.e., when $s$ and $t$ are both on the same domain such as time, it is required that $\C(s,t)=0$ for any $s>t$, so that the future dynamics of $\x(s)$ is not used in the modeling of the current or past dynamics of $\y(t)$. Another interesting direction is to consider sparse and low-rank estimation. For example, to enable the selection of the functional predictors, we could assume that $\V_0$ is a row-sparse matrix and utilize group-wise regularization such as group lasso in estimation. On the theoretical side, it is pressing to study the non-asymptotic behavior of our proposed estimator under reasonable conditions on the integrated design matrix originated from the functional setup. Last but not the least, we will further explore the nested reduced-rank structure, or even more generally, a multi-resolution reduced-rank structure in other statistical problems such as time series analysis and large-scale matrix denoising/approximation tasks.

\section*{Acknowledgment}

Ma's research was partially supported by U.S. NSF grant DMS-1712558. Chen's research was partially supported by U.S. NSF grants DMS-1613295 and IIS-1718798.

\section*{Appendix}
\subsection*{A: Proofs of Main Theoretical Results}

\subsubsection*{Parameterization}

Denote $\Omega$ as the parameter space of the set of matrices $\C \in \mathbb{R}^{J_xp\times J_yd}$ with a nested reduced-rank structure $(\I_{J_x} \otimes \V)\B\A\trans(\I_{J_y}\otimes \U\trans)$ with rank values $(r_x, r_y, r)$. This decomposition is not unique, e.g., with any comfortable and invertible matrices $\Q_1,\Q_2$ and $\Q_3$, we can write
\begin{align}
  \C = & (\I_{J_x} \otimes \V)\B\A\trans(\I_{J_y}\otimes \U\trans) \notag\\
  = & [\I_{J_x} \otimes (\V\Q_1^{-1})][(\I_{J_x} \otimes \Q_1)\B \Q_2\trans][ \Q_2\ntrans \A\trans (\I_{J_y}\otimes \Q_3\trans)] [\I_{J_y}\otimes (\Q_3^{-T}\U\trans)].\label{asy:eq0}
% = & (\I_{J_x} \otimes \V_1)\B_1\A_1\trans(\I_{J_y}\otimes \U_1\trans), 
\end{align}
We therefore consider a reparameterization of $\C$ in order to make its components identifiable and then characterize $\Omega$. Recall that $\V \in \mathbb{R}^{p \times r_x}$ is designed to capture the global low-dimensional structure in predictors and has rank $r_x$ ($\leq p$), then there must exists an invertible sub-matrix $\V_{L_1,\cdot} \in \mathbb{R}^{r_x \times r_x}$ which consists of a set of linearly independent rows. Here $L_1=\{l_1,l_2,\ldots,l_{r_x}\}$ is the row index set. Take $\Q_1=\V_{L_1,\cdot}$ in (\ref{asy:eq0}) and we have $(\V \Q_1^{-1})_{L_1,\cdot}=\I_{r_x}$. Similarly, for $\U$ we can let $\Q_3=\U_{L_3,\cdot} \in \mathbb{R}^{r_y \times r_y}$ such that $(\U \Q_3^{-1})_{L_3,\cdot}=\I_{r_y}$ where $L_3$ is the required row index set.  
%Since $rank(\A)=r$, we can find an invertible sub-matrix $\A_{L,\cdot} \in \mathbb{R}^{r\times r}$ to make $(\A\A_{L,\cdot}^{-1})_{L,\cdot}=\I_{r}$. Then if we want to keep this identity sub-matrix with row index set $L$ in the decomposition, it becomes unique.
Now consider the term $(\I_{J_x} \otimes \Q_1)\B\A\trans(\I_{J_y} \otimes \Q_3\trans) \in \mathbb{R}^{J_x r_x \times J_y r_y}$, which has rank $r$; we can find an invertible sub-matrix $\Q_2 \in \mathbb{R}^{r \times r}$ in $(\I_{J_y} \otimes \Q_3)\A$ to make $((\I_{J_y} \otimes \Q_3)\A \Q_2^{-1})_{L_2,\cdot}=\I_r$. This shows that a nested low-rank matrix can always be reparameterized such that each of $\V$, $\U$, $\A$ is embedded with an identity sub-matrix. With such a representation, $\Omega$ admits a manifold structure that is a union of ${p\choose r_x} \times {d \choose r_y} \times {J_y r_y \choose r}$ many components, i.e.,
\begin{align*}
	\Omega= \cup _{L_1 \times L_2 \times L_3 \in \Pi} \Omega_{L_1 \times L_2 \times L_3}
\end{align*}
where 
\begin{align*}
	\Omega_{L_1 \times L_2 \times L_3}= & \{ (\I_{J_x} \otimes \check\V) \check\B\check\A\trans (\I_{J_y} \otimes \check\U)\trans: \check\V \in \mathbb{R}^{p \times r_x} \ \text{with}\ \check\V_{L_1,\cdot}=\I_{r_x}; \check\B \in \mathbb{R}^{(J_x r_x) \times r }; \\
	&  \check\A \in \mathbb{R}^{(J_y r_y) \times r } \ \text{with} \ \check\A_{L_2,\cdot}=\I_r; \check\U \in \mathbb{R}^{d \times r_y} \ \text{with}\ \check\U_{L_3,\cdot}=\I_{r_y}  \}
\end{align*}
and $\Pi$ consists of all possible index sets $L_1 \times L_2 \times L_3$ with $L_1 \subseteq \{1, \ldots, p\}, L_2 \subseteq\{1, \ldots, J_y r_y\}$ and $L_3 \subseteq \{1, \ldots, d\}$.

% This is equivalent to assuming $\Q_1^*=\V_{0L_1^*,\cdot}$, $\Q_3^*=\U_{0L_3^*,\cdot}$ and $\Q_2^*=((\I_{J_y} \otimes \Q_3^*) \A_0)_{L_2^*,\cdot}$ are three invertible matrices. For simplicity, we use $L_1,L_2,L_3$ and $\Q_1,\Q_2,\Q_3$ to represent $L_1^*,L_2^*,L_3^*$ and $\Q_1^*,\Q_2^*,\Q_3^*$ in the following analysis. Next,

\subsubsection*{Proof of Theorem 1}

\begin{proof}

Based on the above characterization of $\Omega$, we now construct a local neighborhood around the true coefficient matrix $\C_0$, in order to investigate the asymptotic behavior of the NRRR estimation. Suppose $\C_0 \in \Omega_{L_1 \times L_2 \times L_3}$, where $L_1, L_2$ and $L_3$ are three fixed index sets. Define $\check \V=\V_0 \Q_1^{-1}$, $\check \U = \U_0 \Q_3^{-1}$, $\check \B = (\I_{J_x} \otimes \Q_1) \B_0 \Q_2\trans$ and $\check \A = (\I_{J_y} \otimes \Q_3) \A_0 \Q_2^{-1}$ so that $\check \V_{L_1,\cdot}=\I_{r_x}$, $\check \A_{L_2, \cdot}= \I_r$ and $\check \U_{L_3,\cdot}=\I_{r_y}$. It can be verified that
\begin{align*}
	\C_0 & = (\I_{J_x} \otimes \V_0) \B_0 \A_0\trans (\I_{J_y} \otimes \U_0\trans) \\
	     & = [\I_{J_x} \otimes (\V_0 \Q_1^{-1})][ (\I_{J_x} \otimes \Q_1) \B_0 \Q_2\trans][ \Q_2\ntrans \A_0\trans (\I_{J_y} \otimes \Q_3\trans) ][\I_{J_y} \otimes (\Q_3\ntrans \U_0\trans)] \\
	     & = (\I_{J_x} \otimes \check\V) \check\B \check\A\trans (\I_{J_y} \otimes \check\U\trans).
\end{align*}
A local neighborhood centered at $\C_0$ of radius $h > 0$ is constructed as follows,
\begin{align*}
	\mathcal{N}(\C_0, h) = & \bigg\{ \left[\I_{J_x} \otimes \left(\check\V + \frac{1}{\sqrt{n}} \check \R^1 \right)\right] \left(\check\B + \frac{1}{\sqrt{n}} \check \R^2\right)\left(\check\A+ \frac{1}{\sqrt{n}} \check \R^3\right)\trans \left[\I_{J_y} \otimes \left(\check\U+ \frac{1}{\sqrt{n}} \check \R^4\right)\right]\trans: \bigg. \\
	&  \check \R^1 \in \mathbb{R}^{p \times r_x} \ \text{with}\ \check \R^1_{L_1,\cdot}=\0, \|\check \R^1\|_F \leq h; \\
	&  \check \R^2 \in \mathbb{R}^{(J_x r_x) \times r }, \|\check \R^2\|_F \leq h; \\
	&  \check \R^3 \in \mathbb{R}^{(J_y r_y) \times r } \ \text{with} \ \check \R^3_{L_2,\cdot}=\0, \|\check \R^3\|_F \leq h; \\
	&  \bigg. \check \R^4 \in \mathbb{R}^{d \times r_y} \ \text{with}\ \check \R^4_{L_3,\cdot}=\0, \|\check \R^4\|_F \leq h \bigg\}.
\end{align*}
The zero parts in perturbation matrices $\check \R^1$, $\check \R^3$ and $\check \R^4$ ensure that $\mathcal{N}(\C_0, h) \subseteq \Omega_{L_1 \times L_2 \times L_3} \subseteq \mathcal{N}(\C_0, \infty) $. Also note that we can equivalently express the neighborhood in terms of $\V_0,\B_0,\A_0,\U_0$ as
\begin{align*}
	\mathcal{N}(\C_0, h) = & \bigg\{ \left[\I_{J_x} \otimes \left(\V_0 + \frac{1}{\sqrt{n}} \R^1 \right)\right] \left(\B_0 + \frac{1}{\sqrt{n}} \R^2\right)\left(\A_0+ \frac{1}{\sqrt{n}} \R^3\right)\trans \left[\I_{J_y} \otimes \left(\U_0+ \frac{1}{\sqrt{n}} \R^4\right)\right]\trans: \bigg. \\
	&  \R^1 = \check \R^1\Q_1 \in \mathbb{R}^{p \times r_x} \ \text{with}\ \R^1_{L_1,\cdot}=\check \R^1_{L_1,\cdot}=\0, \|\check \R^1\|_F \leq h; \\
	&  \R^2 = (\I_{J_x} \otimes \Q_1^{-1})\check \R^2\Q_2\trans \in \mathbb{R}^{(J_x r_x) \times r }, \|\check \R^2\|_F \leq h; \\
	&  \R^3 = (\I_{J_y} \otimes \Q_3^{-1}) \check \R^3 \Q_2 \in \mathbb{R}^{(J_y r_y) \times r } \ \text{with} \ \check \R^3_{L_2,\cdot}=\0, \|\check \R^3\|_F \leq h; \\
	& \bigg. \R^4 = \check \R^4 \Q_3 \in \mathbb{R}^{d \times r_y} \ \text{with}\ \R^4_{L_3,\cdot} = \check \R^4_{L_3,\cdot}=\0, \|\check \R^4\|_F \leq h \bigg\}.
\end{align*}

With the above setup, we now investigate the consistency of the NRRR estimation that minimizes the objective function 
\begin{align*}
	\Q_n(\V,\B,\A,\U)=\|\Y - \X (\I_{J_x} \otimes \V) \B \A\trans (\I_{J_y} \otimes \U\trans) \|_F^2.
\end{align*}
We claim that for any given $\epsilon > 0$, there exists a large enough constant $h$ such that 
	\small
	\begin{align}\label{asy:eq1}
		\mathbb{P}\bigg\{ \inf_{\|\check \R^1\|_F=\|\check \R^2\|_F=\|\check \R^3\|_F=\|\check \R^4\|_F=h} & \Q_n(\V_0 + \frac{1}{\sqrt{n}} \R^1, \B_0 + \frac{1}{\sqrt{n}} \R^2, \A_0+ \frac{1}{\sqrt{n}} \R^3, \U_0+ \frac{1}{\sqrt{n}} \R^4) \bigg. \nonumber\\ 
	&  \bigg. > \Q_n(\V_0,\B_0,\A_0,\U_0)  \bigg\} \geq 1-\epsilon.
	\end{align}
	\normalsize
This statement implies that with probability at least $1-\epsilon$, there exists a local minimum $\widehat \C=(\I_{J_x} \otimes \widehat\V) \widehat\B \widehat\A\trans (\I_{J_y} \otimes \widehat\U\trans)$ in the interior of the ball $\mathcal{N}(\C_0,h)$ and it satisfies 
	\small
	\begin{align*}
		& \|(\widehat \V -\V_0) \Q_1^{-1}\|_F = O_p(n^{-\frac{1}{2}}), \\
		& \|(\I_{J_x} \otimes \Q_1)(\widehat \B -\B_0) \Q_2\ntrans \|_F = O_p(n^{-\frac{1}{2}}), \\
		& \|(\I_{J_y} \otimes \Q_3)(\widehat \A -\A_0) \Q_2^{-1}\|_F = O_p(n^{-\frac{1}{2}}), \\
		& \|(\widehat \U -\U_0) \Q_3^{-1}\|_F = O_p(n^{-\frac{1}{2}}). 
	\end{align*}
	\normalsize
	Then based on the fact $\|\A\B\|_F \leq \|\A\|_F\|\B\|_F$ for any two matrices $\A$ and $\B$, we have 
	\begin{align*}
		\|(\widehat \V - \V_0) \Q_1^{-1} \|_F\|\Q_1^{-1} \|_F^{-1} \leq \| \widehat \V - \V_0 \|_F \leq \|(\widehat \V - \V_0) \Q_1^{-1} \|_F\|\Q_1\|_F.
	\end{align*}
	Thus with $\|\Q_1\|_F < \infty$ we obtain $\| \widehat \V - \V_0 \|_F=O_p(n^{-\frac{1}{2}})$. Similarly, with $\| \Q_2 \|_F < \infty$ and $\|\Q_3\|_F < \infty$ we can obtain $\|\widehat \B - \B_0 \|_F=O_p(n^{-\frac{1}{2}})$, $\|\widehat \A - \A_0 \|_F=O_p(n^{-\frac{1}{2}})$ and $\|\widehat \U - \U_0 \|_F=O_p(n^{-\frac{1}{2}})$.

It remains to verify (\ref{asy:eq1}). Let's write 
\small
\begin{align*}
	\left[\I_{J_x} \otimes \left(\V_0 + \frac{1}{\sqrt{n}} \R^1 \right)\right] \left(\B_0 + \frac{1}{\sqrt{n}} \R^2\right)\left(\A_0+ \frac{1}{\sqrt{n}} \R^3\right)\trans \left[\I_{J_y} \otimes \left(\U_0+ \frac{1}{\sqrt{n}} \R^4\right)\right]\trans \in \mathcal{N}(\C_0,h)
\end{align*}
\normalsize
as any perturbed matrix within $\mathcal{N}(\C_0,h)$ and define 
\begin{align*}
	\Psi_n(\R^1,\R^2,\R^3,\R^4)= & \Q_n(\V_0 + n^{-\frac{1}{2}} \R^1, \B_0 + n^{-\frac{1}{2}} \R^2, \A_0+ n^{-\frac{1}{2}} \R^3, \U_0+ n^{-\frac{1}{2}} \R^4) \\
	& -\Q_n(\V_0,\B_0,\A_0,\U_0).
\end{align*}
By some algebra, we get 
\begin{align}\label{asy:eq2}
	\Psi_n(\R^1,\R^2,\R^3,\R^4)=-\Z\trans \mbox{vec}\left(\frac{\X\trans \E}{\sqrt{n}}\right)+ \Z\trans\left(\I_{J_y d} \otimes \frac{\X\trans\X}{n}\right)\Z + O_p(n^{-\frac{1}{2}})
\end{align}
where 
\begin{align*}
	\Z = & vec((\I_{J_x} \otimes \R^1) \B_0 \A_0\trans (\I_{J_y} \otimes \U_0\trans)+(\I_{J_x} \otimes \V_0) \R^2 \A_0\trans (\I_{J_y} \otimes \U_0\trans) \\
	      & +(\I_{J_x} \otimes \V_0) \B_0 \R^{3\rm T} (\I_{J_y} \otimes \U_0\trans)+(\I_{J_x} \otimes \V_0) \B_0 \A_0\trans (\I_{J_y} \otimes \R^{4\rm T})).
\end{align*}
%Now it suffices to show that for a sufficiently large $h$, the second term on the right-hand side of (\ref{asy:eq2}) dominates the first term for $(R^1,R^2,R^3,R^4)$ with $\|\check R^1\|_F=\|\check R^2\|_F=\|\check R^3\|_F=\|\check R^4\|_F=h$. Furthermore, 
%From Assumption 2 we also have 
%$$
%vec\left(\frac{1}{\sqrt{n}} \X\trans\E \right) \rightarrow^d N(0, \bSig \otimes \bG).  
%$$       

Because $$
\mbox{vec}\left(\frac{1}{\sqrt{n}} \X\trans\E \right) \rightarrow^d N(\0, \bSig \otimes \bG)  
$$ and $$ \I_{J_y d} \otimes \frac{\X\trans\X}{n} \rightarrow \I_{J_y d} \otimes \bG,$$
it suffices to show that for a large enough $h$, denoted as $h_n^*$, $\|\Z\|^2$ dominates $\|\Z\|$ for $(\R^1,\R^2,\R^3,\R^4)$ with $\|\check \R^1\|_F=\|\check \R^2\|_F=\|\check \R^3\|_F=\|\check \R^4\|_F=h$. For simplicity, write $\Z=\mbox{vec}(\P_1+\P_1+\P_3+\P_4)$ where 
	\begin{align*}
		& \P_1=(\I_{J_x} \otimes \R^1) \B_0 \A_0\trans (\I_{J_y} \otimes \U_0\trans),\\
		& \P_2=(\I_{J_x} \otimes \V_0) \R^2 \A_0\trans (\I_{J_y} \otimes \U_0\trans), \\
	    & \P_3=(\I_{J_x} \otimes \V_0) \B_0 \R^{3 \rm T} (\I_{J_y} \otimes \U_0\trans),\\
	    & \P_4=(\I_{J_x} \otimes \V_0) \B_0 \A_0\trans (\I_{J_y} \otimes \R^{4\rm T}),
	\end{align*}
	and also write 
	\begin{align*}
		& \B_0=( \B_{01}\trans, \cdots, \B_{0 J_x}\trans )\trans, \quad \B_{0i} \in \mathbb{R}^{r_x \times r}, \quad i=1,\ldots,J_x \\
		& \A_0=( \A_{01}\trans, \cdots, \A_{0 J_y}\trans )\trans, \quad \A_{0j} \in \mathbb{R}^{r_y \times r}, \quad j=1,\ldots,J_y \\
		& \check \R^2=( \check \R^{2\rm T}_1, \ldots, \check \R^{2\rm T}_{J_x} )\trans, \quad \check \R^2_i \in \mathbb{R}^{r_x \times r}, \quad i=1,\ldots,J_x \\
		& \check \R^3=( \check \R^{3\rm T}_1, \ldots, \check \R^{3\rm T}_{J_y} )\trans, \quad \check \R^3_j \in \mathbb{R}^{r_y \times r}, \quad j=1,\ldots,J_y. 
	\end{align*}		
	%The following analysis will proceed sequentially in the order of $\P_1,\P_4$ and $\P_2, \P_3$.
Let's first consider 
	$$
	\P_1=\begin{pmatrix} 
     \R^1\B_{01}\A\trans_{01}\U_0\trans & \cdots & \R^1\B_{01}\A\trans_{0J_y}\U_0\trans \\
     \vdots & \ddots & \vdots \\
     \R^1\B_{0J_x}\A\trans_{01}\U_0\trans & \cdots & \R^1\B_{0J_x}\A\trans_{0J_y}\U_0\trans
    \end{pmatrix},
	$$
	where each block $\R^1\B_{0i}\A\trans_{0j}\U_0\trans \in \mathbb{R}^{p \times q}$ for $i=1,\ldots,J_x,\ j=1,\ldots, J_y$. Recall that $\R^1_{L_1,\cdot}=\0$, thus we have $(\R^1\B_{0i}\A\trans_{0j}\U_0\trans)_{L_1,\cdot}=\0$. Without loss of generality, we assume $L_1=\{1, \ldots, r_x\}$. Then, if we write $\P_1=(\P_{11}\trans,\ldots,\P_{1J_x}\trans)\trans$ with $\P_{1i} \in \mathbb{R}^{p \times (J_y d)}, i=1,\ldots, J_x$, the first $r_x$ rows for each $\P_{1i}$ are zero vectors. 
	
	Next we deal with $\P_4$ which can be written as
		$$
	\P_4=\begin{pmatrix} 
     \V_0\B_{01}\A\trans_{01}\R^{4\rm T} & \cdots & \V_0\B_{01}\A\trans_{0J_y}\R^{4\rm T} \\
     \vdots & \ddots & \vdots \\
     \V_0\B_{0J_x}\A\trans_{01}\R^{4\rm T} & \cdots & \V_0\B_{0J_x}\A\trans_{0J_y}\R^{4\rm T}
    \end{pmatrix},
	$$
	where each block $\V_0\B_{0i}\A\trans_{0j}\R^{4\rm T} \in \mathbb{R}^{p \times q}$ for $i=1,\ldots,J_x,\ j=1,\ldots, J_y$. With $\R^4_{L_3,\cdot}=\0$ we have $(\V_0\B_{0i}\A\trans_{0j}R^{4\rm T})_{\cdot,L_3}=\0$. Without loss of generality, we assume $L_3=\{1, \ldots, r_y\}$. If we write $\P_4=(\P_{41},\ldots,\P_{4J_y})$ with $\P_{4j} \in \mathbb{R}^{(J_x p)\times d}, j=1,\ldots, J_y$, the first $r_y$ columns for each $\P_{4j}$ are zero vectors. To summarize, for each block $\R^1\B_{0i}\A\trans_{0j}\U_0\trans+\V_0\B_{0i}\A\trans_{0j}\R^{4\rm T}$ in $\P_1+\P_4$, the left-upper sub-matrix is a zero matrix of dimension $r_x \times r_y$.

	Then we consider  
	\begin{align*}
		\P_2 & = (\I_{J_x} \otimes \V_0) \R^2 \A_0\trans (\I_{J_y} \otimes \U_0\trans) \\
		     & = (\I_{J_x} \otimes \check\V) \check \R^2 \check\A\trans (\I_{J_y} \otimes \check\U\trans) \\
		     & =		
	\begin{pmatrix} 
     \check\V\check \R^2_{1}\check\A\trans_{1}\check\U\trans & \cdots & \check\V\check \R^2_{1}\check\A\trans_{J_y}\check\U\trans \\
     \vdots & \ddots & \vdots \\
     \check\V\check \R^2_{J_x}\check\A\trans_{1}\check\U\trans & \cdots & \check\V\check \R^2_{J_x}\check\A\trans_{J_y}\check\U\trans
    \end{pmatrix}.
	\end{align*}
	For each block $\check\V\check \R^2_{i}\check\A\trans_{j}\check\U\trans \in \mathbb{R}^{p \times d}$, we have $(\check\V\check \R^2_{i}\check\A\trans_{j}\check\U\trans)_{L_1,L_3} = \check \R^2_{i}\check\A\trans_{j} \in \mathbb{R}^{r_x \times r_y}$ because $\check \V_{L_1,\cdot}=\I_{r_x}$ and $\check \U_{L_3,\cdot}=\I_{r_y}$. Similarly, we have
\begin{align*}
		\P_3 & =(\I_{J_x} \otimes \check\V) \check \B \check \R^{3\rm T} (\I_{J_y} \otimes \check\U\trans) \\
		     & = 		
	\begin{pmatrix} 
     \check\V\check \B_{1}\check \R^{3\rm T}_{1}\check\U\trans & \cdots & \check\V\check \B_{1}\check \R^{3\rm T}_{J_y}\check\U\trans \\
     \vdots & \ddots & \vdots \\
     \check\V\check \B_{J_x}\check \R^{3\rm T}_{1}\check\U\trans & \cdots & \check\V\check \B_{J_x}\check \R^{3\rm T}_{J_y}\check\U\trans
    \end{pmatrix}
	\end{align*}
	where each block $\check\V\check \B_{i}\check \R^{3\rm T}_{j}\check\U\trans \in \mathbb{R}^{p \times d}$ and $(\check\V\check \B_{i}\check \R^{3\rm T}_{j}\check\U\trans)_{L_1,L_3} = \check \B_{i}\check \R^{3\rm T}_{j} \in \mathbb{R}^{r_x \times r_y}$. % because $\check \U_{L_1,\cdot}=\I_{r_x}$ and $\check \V_{L_3,\cdot}=\I_{r_y}$. 
	Thus, if we extract the left upper sub-matrix which has dimension $r_x \times r_y$ from all blocks in $\P_1+\P_2+\P_3+\P_4$ and put them together, we can obtain a matrix 
	\begin{align*}
		\begin{pmatrix} 
     \check \R^2_{1}\check\A\trans_{1}+\check \B_{1}\check \R^{3\rm T}_{1} & \cdots & \check \R^2_{1}\check\A\trans_{J_y}+\check \B_{1}\check \R^{3\rm T}_{J_y} \\
     \vdots & \ddots & \vdots \\
     \check \R^2_{J_x}\check\A\trans_{1}+\check \B_{J_x}\check \R^{3\rm T}_{1} & \cdots & \check \R^2_{J_x}\check\A\trans_{J_y}+\check \B_{J_x}\check R^{3\rm T}_{J_y}
    \end{pmatrix}
    =	\check \R^2 \check \A\trans + \check \B \check \R^{3\rm T}.	
	\end{align*}
	From $\check \R^3_{L_2,\cdot}=\0$, we have $(\check \B\check \R^{3\rm T})_{\cdot,L_2}=\0$. And from $\check \A_{L_2,\cdot}=\I_r$, we have $(\check \R^2 \check \A\trans)_{\cdot,L_2}=\check \R^2$. It leads to $(\check \B\check \R^{3\rm T} + \check \R^2 \check \A\trans)_{\cdot,L_2}=\check \R^2$ and 
	$\| (\check \B\check \R^{3\rm T} + \check \R^2 \check \A\trans)_{\cdot,L_2} \|_F^2 = \|\check \R^2 \|_F^2 = h^2 $. Recall that $(\check \B\check \R^{3\rm T} + \check \R^2 \check \A\trans)_{\cdot,L_2}$ is a sub-matrix in $\P_1+\P_2+\P_3+\P_4$. Then with %a similar statement in \citet{chen2012reduced}, i.e., 
	\begin{align*}
		\|\Z\|^2=\| \P_1+\P_2+\P_3+\P_4 \|_F^2= h^2 \left(1+f \left( \frac{\check R^1}{h}, \frac{\check R^2}{h}, \frac{\check R^3}{h}, \frac{\check R^4}{h} \right)\right) \geq (k+1) h^2,
	\end{align*}
	where $f()$ is a non-negative, continuous function which attains its minimum value $k>0$ over the unit sphere $\{ (\check \R^1, \check \R^2, \check \R^3, \check \R^4): \| \check \R^1 \|_F=1, \| \check \R^2 \|_F=1, \| \check \R^3 \|_F=1, \| \check \R^4 \|_F=1; \check \R^1_{L_1, \cdot}=\0, \check \R^3_{L_2, \cdot}=\0,  \check \R^4_{L_3, \cdot}=\0\}$. We have verified the existence of $h_n^*$ due to the fact that $\|\Z\|$ is $O_p(h)$ uniformly. This completes the proof. %Moreover, due to the law of iterated logarithm, we can choose $h_n^*=O(\sqrt{\log(\log(n))})$ to enlarge the neighborhood and make the local minimizer to be a global minimizer.
\end{proof}

\subsubsection*{Proof of Theorem 2}

\begin{proof}
By the definition of $\widehat \C$, we have
\begin{equation*}
\| \Y-\X \widehat \C \| _{F}^{2}\leq
\| \Y-\mathbf{XC}_{0} \| _{F}^{2},
\end{equation*}%
which leads to 
\begin{equation}
\| \X ( \widehat \C-\C_0) \trans \|
_F^2 \leq 2 \langle \E,\X ( \widehat\C-\C_0)\trans\rangle _F,  \label{EQ:Chat-C}
\end{equation}%
where $\langle \C,\D \rangle _F = \text{tr}(\C\trans \D)$. Furthermore,
\begin{equation}
\langle \E,\X( \widehat \C-\C_0)\trans \rangle _F=\langle \mathcal{P}\E,\X (
\widehat \C-\C_0) \trans \rangle _F,
\label{EQ:PE-C}
\end{equation}%
where $\mathcal{P} = \X\left( \X\trans \X \right) ^{-}\X\trans$ denotes the projection matrix onto the column space of $\X$. Let $d_j( \cdot)$ denote the $j^{\mbox{th}}$ largest singular value of the enclosed matrix. Then we have $\langle \C,\D \rangle _F \leq d_1(\C) \| \D \|_*$, where $\| \D \|_* =\sum\nolimits_j d_j ( \D ) $ denotes the nuclear norm of $\D$. It follows that 
\begin{eqnarray}
\langle \mathcal{P}\E,\X( \widehat \C-\C_0) \trans \rangle _F &\leq &d_1( \mathcal{P}\E) \| \X( \widehat \C-\C_0) \trans \|_* \notag \\
&\leq &d_1 ( \mathcal{P}\E) \sqrt{2r} \| \X (\widehat \C-\C_0) \trans \| _F.  \label{EQ:PE}
\end{eqnarray}%
Therefore, by (\ref{EQ:Chat-C}) (\ref{EQ:PE-C}), and (\ref{EQ:PE}), we have
\begin{equation*}
\| \X ( \widehat \C-\C_0) \trans \|
_F \leq 2 d_1 ( \mathcal{P}\E) \sqrt{2r}.
\end{equation*}%
By Lemma 3 in \citet{bunea2011}, we have $$\mathbb{E}\left\{ d_{1}( \mathcal{P}\E) \right\} \leq \sigma\left( \sqrt{r(\X)}+\sqrt{dJ_y}\right)$$ 
and
$$
\mathbb{P}\left\{d_1(\mathcal{P}\E) \geq \mathbb{E}[d_1(\mathcal{P}\E )] + \sigma \theta \left(\sqrt{r(\X)} + \sqrt{dJ_y}\right)  \right\} \leq \exp\left\{-\frac{\theta^2}{2}(r(\X)+ dJ_y)\right\}
$$
where $\theta$ is a positive constant. Therefore, $\| \X( \widehat \C-\C_0) \trans\| _F=O_p \left( r^{1/2} (r(\X) + dJ_y)^{1/2} \right)$ with probability at least $1-\exp\left\{-\theta^2(r(\X)+ dJ_y)/2\right\}$. The second result follows directly.

%By Assumption (A1), we have $\| \widehat \C-\C_0 \| _F=O_{p}\left( \sqrt{J_n r/n} \right) $. 
\end{proof}

\subsection*{B: Additional Simulation Results}
\subsubsection*{Additional Simulation Results from BIC Tuning}
We present the results on estimating $r$, $r_x$ and $r_y$ in this part. For methods with nested reduced-rank structure, BIC is exploited to select ranks.

\begin{table}[htbp]
  \centering
  \caption{Simulation results for Setting 1 with true rank $r=5$. The mean and the percentage of matching with the true rank (in parenthesis) of rank estimation over 300 simulation runs are presented.}\label{tab:rank1}
  \begin{tabular}{lrrrrrr}
    \hline
     & \multicolumn{1}{c}{$\rho$} & \multicolumn{1}{c}{NRRR} & \multicolumn{1}{c}{NRRR-X} & \multicolumn{1}{c}{RRR} & \multicolumn{1}{c}{RRS} & \multicolumn{1}{c}{NRRS}\\\hline
    \hfill  & 0.1       & 3.58 (0.09) & 2.46 (0.00) & 1.80 (0.00) & 4.15 (0.28) & 4.02 (0.22) \\
    SNR=1 \ & 0.5 & 3.17 (0.02) & 2.24 (0.00) & 1.65 (0.00) & 3.95 (0.19) & 3.51 (0.06) \\
    \hfill & 0.9        & 2.13 (0.00) & 1.66 (0.00) & 1.55 (0.00) & 3.30 (0.03) & 2.22 (0.00) \\\hline
    \hfill & 0.1        & 4.88 (0.88) & 4.37 (0.44) & 3.65 (0.11) & 4.91 (0.91) & 4.91 (0.91) \\
    SNR=2 \ & 0.5 & 4.71 (0.72) & 4.01 (0.19) & 3.40 (0.04) & 4.80 (0.80) & 4.82 (0.82) \\
    \hfill & 0.9        & 3.72 (0.09) & 3.05 (0.00) & 2.69 (0.00) & 4.16 (0.31) & 3.92 (0.16) \\\hline
    \hfill & 0.1        & 5.00 (1.00) & 4.97 (0.97) & 4.88 (0.89) & 4.99 (0.99) & 5.00 (1.00) \\
    SNR=4 \ & 0.5 & 5.00 (1.00) & 4.90 (0.90) & 4.66 (0.68) & 4.99 (0.99) & 5.00 (1.00) \\
    \hfill & 0.9        & 4.74 (0.73) & 4.22 (0.33) & 3.72 (0.12) & 4.83 (0.84) & 4.73 (0.74) \\
    \hline
  \end{tabular}
\end{table}

\begin{table}[htbp]
  \centering
  \caption{Simulation results for Setting 2 with true rank $r=3$. The mean and the percentage of matching with the true rank (in parenthesis) of rank estimation over 300 simulation runs are presented.}\label{tab:rank2}
  \begin{tabular}{lrrrrrr}
    \hline
     & \multicolumn{1}{c}{$\rho$} & \multicolumn{1}{c}{NRRR} & \multicolumn{1}{c}{NRRR-X} & \multicolumn{1}{c}{RRR} & \multicolumn{1}{c}{RRS} & \multicolumn{1}{c}{NRRS}\\\hline
    \hfill & 0.1        & 2.99 (0.99) & 2.37 (0.40) & 2.31 (0.58) & 2.91 (0.91) & 2.99 (0.99) \\
    SNR=1 \ & 0.5 & 2.94 (0.94) & 2.20 (0.29) & 2.33 (0.47) & 2.83 (0.83) & 2.98 (0.98) \\
    \hfill & 0.9        & 2.51 (0.54) & 1.89 (0.08) & 2.12 (0.21) & 2.64 (0.64) & 2.71 (0.73) \\\hline
    \hfill & 0.1        & 3.00 (1.00) & 2.97 (0.97) & 2.62 (0.87) & 3.09 (0.94) & 3.00 (1.00) \\
    SNR=2 \ & 0.5 & 3.00 (1.00) & 2.93 (0.93) & 2.77 (0.89) & 3.02 (0.92) & 3.00 (1.00) \\
    \hfill & 0.9        & 2.98 (0.98) & 2.72 (0.72) & 2.77 (0.77) & 3.02 (0.96) & 3.00 (1.00) \\\hline
    \hfill & 0.1        & 3.00 (1.00) & 3.00 (1.00) & 2.67 (0.89) & 3.13 (0.93) & 3.00 (1.00) \\
    SNR=4 \ & 0.5 & 3.00 (1.00) & 3.00 (1.00) & 2.81 (0.93) & 3.15 (0.95) & 3.00 (1.00) \\
    \hfill & 0.9        & 3.00 (1.00) & 2.97 (0.97) & 2.96 (0.96) & 3.00 (0.98) & 3.00 (1.00) \\
    \hline
  \end{tabular}
\end{table}

\begin{table}[htbp]
  \centering
  \caption{Simulation results for Setting 1 with true rank $r_x=3$. The mean and the percentage of matching with the true rank (in parenthesis) of $r_x$ estimation over 300 simulation runs are presented. }\label{tab:rx1}
  \begin{tabular}{lrrrr}
    \hline
    & \multicolumn{1}{c}{$\rho$} & \multicolumn{1}{c}{NRRR} & \multicolumn{1}{c}{NRRR-X} & \multicolumn{1}{c}{NRRS}\\\hline
    \hfill & 0.1       & 2.63 (0.64) & 2.62 (0.63) & 2.75 (0.77) \\
    SNR=1 \ & 0.5 & 2.41 (0.49) & 2.40 (0.47) & 2.50 (0.54) \\
    \hfill & 0.9       & 1.75 (0.13) & 1.77 (0.13) & 1.85 (0.19) \\\hline
    \hfill & 0.1        & 3.00 (1.00) & 3.00 (1.00) & 3.00 (1.00) \\
    SNR=2 \ & 0.5 & 3.00 (1.00) & 3.00 (1.00) & 3.00 (1.00) \\
    \hfill & 0.9        & 2.88 (0.87) & 2.85 (0.84) & 2.90 (0.90) \\\hline
    \hfill & 0.1        & 3.00 (1.00) & 3.00 (1.00) & 3.00 (1.00) \\
    SNR=4 \ & 0.5 & 3.00 (1.00) & 3.00 (1.00) & 3.00 (1.00) \\
    \hfill & 0.9        & 3.01 (0.99) & 3.01 (0.99) & 3.00 (1.00) \\
    \hline
  \end{tabular}
\end{table}

\begin{table}[htbp]
  \centering
  \caption{Simulation results for Setting 2 with true rank $r_x=3$. The mean and the percentage of matching with the true rank (in parenthesis) of $r_x$ estimation over 300 simulation runs are presented. }\label{tab:rx2}
  \begin{tabular}{lrrrr}
    \hline
    & \multicolumn{1}{c}{$\rho$} & \multicolumn{1}{c}{NRRR} & \multicolumn{1}{c}{NRRR-X} & \multicolumn{1}{c}{NRRS}\\\hline
    \hfill & 0.1        & 2.93 (0.93) & 2.90 (0.90) & 2.98 (0.88) \\
    SNR=1 \ & 0.5 & 2.88 (0.89) & 2.87 (0.88) & 2.92 (0.91) \\
    \hfill & 0.9        & 2.44 (0.49) & 2.44 (0.49) & 2.55 (0.59) \\\hline
    \hfill & 0.1        & 2.94 (0.94) & 2.94 (0.94) & 3.24 (0.78) \\
    SNR=2 \ & 0.5 & 2.94 (0.94) & 2.96 (0.96) & 3.14 (0.86) \\
    \hfill & 0.9        & 2.98 (0.96) & 2.99 (0.97) & 3.09 (0.89) \\\hline
    \hfill & 0.1        & 2.97 (0.96) & 2.98 (0.98) & 3.62 (0.85) \\
    SNR=4 \ & 0.5 & 2.97 (0.97) & 2.96 (0.97) & 3.21 (0.93) \\
    \hfill & 0.9        & 3.00 (1.00) & 3.00 (1.00) & 3.09 (0.93) \\
    \hline
  \end{tabular}
\end{table}

\begin{table}[htbp]
  \centering
  \caption{Simulation results for Setting 1 with true rank $r_y=3$. The mean and the percentage of matching with the true rank (in parenthesis) of $r_y$ estimation over 300 simulation runs are presented. }\label{tab:ry1}
  \begin{tabular}{lrrrr}
    \hline
    & \multicolumn{1}{c}{$\rho$} & \multicolumn{1}{c}{NRRR} & \multicolumn{1}{c}{NRRS}\\\hline
    \hfill & 0.1        & 2.92 (0.92) & 2.97 (0.97) \\
    SNR=1 \ & 0.5 & 2.95 (0.95) & 2.99 (0.99) \\
    \hfill & 0.9        & 2.95 (0.95) & 2.98 (0.98) \\\hline
    \hfill & 0.1        & 3.00 (1.00) & 3.00 (1.00) \\
    SNR=2 \ & 0.5 & 3.00 (1.00) & 3.00 (1.00) \\
    \hfill & 0.9        & 3.01 (0.99) & 3.00 (1.00) \\\hline
    \hfill & 0.1        & 3.00 (1.00) & 3.00 (1.00) \\
    SNR=4 \ & 0.5 & 3.00 (1.00) & 3.00 (1.00) \\
    \hfill & 0.9        & 3.00 (1.00) & 3.00 (1.00) \\
    \hline
  \end{tabular}
\end{table}

\begin{table}[htbp]
  \centering
  \caption{Simulation results for Setting 2 with true rank $r_y=3$. The mean and the percentage of matching with the true rank (in parenthesis) of $r_y$ estimation over 300 simulation runs are presented. }\label{tab:ry2}
  \begin{tabular}{lrrrr}
    \hline
    & \multicolumn{1}{c}{$\rho$} & \multicolumn{1}{c}{NRRR} & \multicolumn{1}{c}{NRRS}\\\hline
    \hfill & 0.1        & 2.99 (0.99) & 2.99 (0.99) \\
    SNR=1 \ & 0.5 & 3.00 (1.00) & 3.00 (1.00) \\
    \hfill & 0.9        & 3.00 (0.99) & 3.00 (1.00) \\\hline
    \hfill & 0.1        & 2.99 (0.99) & 3.00 (1.00) \\
    SNR=2 \ & 0.5 & 3.00 (1.00) & 3.00 (1.00) \\
    \hfill & 0.9        & 3.00 (1.00) & 3.00 (1.00) \\\hline
    \hfill & 0.1        & 3.00 (1.00) & 3.00 (1.00) \\
    SNR=4 \ & 0.5 & 3.00 (1.00) & 3.00 (1.00) \\
    \hfill & 0.9        & 3.00 (1.00) & 3.00 (1.00) \\
    \hline
  \end{tabular}
\end{table}

\clearpage
\subsubsection*{Simulation Results from Cross Validation Tuning}
We present simulation results under Setting 1 with all the ranks selected by 10-fold cross validation. Results of MSPE and MSFPE displayed here are the trimmed version with the smallest and the largest 20 observations deleted from 300 simulation runs.

\begin{table}[htbp]
  \centering
  \caption{Simulation results for Setting 1. The trimmed means and standard deviations (in parenthesis) of MSPE are presented. To improve presentation, all values are multiplied by 10.}\label{tab:ypred_cv}
  \begin{tabular}{lrrrrrr}
    \hline
    & \multicolumn{1}{c}{$\rho$}  & \multicolumn{1}{c}{NRRR} & \multicolumn{1}{c}{NRRR-X} & \multicolumn{1}{c}{RRR} & \multicolumn{1}{c}{RRS} & \multicolumn{1}{c}{NRRS}\\\hline
   \hfill      & 0.1 & 11.05 (2.54) & 11.42 (2.61) & 14.67 (3.24) & 11.52 (2.56) & 10.80 (2.46) \\ 
  SNR=1 \ & 0.5 & 17.34 (4.30) & 17.84 (4.37) & 22.40 (5.33) & 17.45 (4.14) & 16.76 (4.06) \\ 
  \hfill        & 0.9 & 26.05 (8.72) & 26.37 (8.74) & 30.28 (9.51) & 24.39 (7.79) & 24.30 (8.03) \\ \hline
  \hfill        & 0.1 & 2.72 (0.51) & 2.81 (0.52) & 3.91 (0.75) & 3.12 (0.55) & 2.81 (0.51) \\ 
  SNR=2 \ & 0.5 & 4.07 (0.91) & 4.21 (0.95) & 5.78 (1.29) & 4.48 (0.97) & 4.12 (0.90) \\ 
  \hfill        & 0.9 & 6.05 (1.98) & 6.23 (2.03) & 8.00 (2.55) & 6.20 (1.96) & 5.89 (1.88) \\  \hline
  \hfill        & 0.1 & 0.66 (0.14) & 0.68 (0.14) & 0.93 (0.20) & 0.96 (0.19) & 0.77 (0.17) \\ 
  SNR=4 \ & 0.5 & 1.04 (0.24) & 1.08 (0.25) & 1.46 (0.34) & 1.31 (0.26) & 1.15 (0.24) \\ 
  \hfill        & 0.9 & 1.50 (0.48) & 1.56 (0.50) & 2.10 (0.66) & 1.67 (0.49) & 1.57 (0.47) \\ 
   \hline
  \end{tabular}
\end{table}

\begin{table}[htbp]
  \centering
  \caption{Simulation results for Setting 1 with true rank $r=5$. The mean and the percentage of matching with the true rank (in parenthesis) of rank estimation over 300 simulation runs are presented. }\label{tab:rank_cv}
  \begin{tabular}{lrrrrrr}
    \hline
    & \multicolumn{1}{c}{$\rho$} & \multicolumn{1}{c}{NRRR} & \multicolumn{1}{c}{NRRR-X} & \multicolumn{1}{c}{RRR} & \multicolumn{1}{c}{RRS} & \multicolumn{1}{c}{NRRS}\\\hline
    \hfill & 0.1        & 4.73 (0.75) & 4.20 (0.36) & 1.75 (0.00) & 4.12 (0.26) & 4.96 (0.85) \\
    SNR=1 \ & 0.5 & 4.36 (0.52) & 3.80 (0.18) & 1.70 (0.00) & 3.96 (0.19) & 4.88 (0.74) \\
    \hfill & 0.9        & 2.81 (0.06) & 2.40 (0.01) & 1.53 (0.00) & 3.29 (0.02) & 4.39 (0.33) \\\hline
    \hfill & 0.1        & 5.00 (1.00) & 4.96 (0.96) & 3.68 (0.12) & 4.91 (0.91) & 5.06 (0.94) \\
    SNR=2 \ & 0.5 & 4.99 (0.98) & 4.91 (0.91) & 3.48 (0.04) & 4.83 (0.83) & 5.07 (0.94) \\
    \hfill & 0.9        & 4.63 (0.62) & 4.17 (0.36) & 2.70 (0.00) & 4.26 (0.38) & 5.02 (0.80) \\\hline
    \hfill & 0.1        & 5.00 (1.00) & 5.00 (1.00) & 4.84 (0.86) & 4.97 (0.97) & 5.03 (0.98) \\
    SNR=4 \ & 0.5 & 5.00 (1.00) & 5.00 (1.00) & 4.76 (0.76) & 5.00 (1.00) & 5.04 (0.96) \\
    \hfill & 0.9        & 5.00 (0.93) & 4.85 (0.85) & 3.70 (0.11) & 4.83 (0.83) & 5.12 (0.89) \\
    \hline
  \end{tabular}
\end{table}

\begin{table}[htbp]
  \centering
  \caption{Simulation results for Setting 1 with true rank $r_x=3$. The mean and the percentage of matching with the true rank (in parenthesis) of $r_x$ estimation over 300 simulation runs are presented. }\label{tab:rx_cv}
  \begin{tabular}{lrrrr}
    \hline
    & \multicolumn{1}{c}{$\rho$} & \multicolumn{1}{c}{NRRR} & \multicolumn{1}{c}{NRRR-X} & \multicolumn{1}{c}{NRRS}\\\hline
    \hfill & 0.1        & 2.85 (0.84) & 2.85 (0.84) & 3.05 (0.82) \\
    SNR=1 \ & 0.5 & 2.67 (0.68) & 2.67 (0.68) & 3.08 (0.72) \\
    \hfill & 0.9        & 1.97 (0.21) & 1.97 (0.21) & 3.22 (0.41) \\\hline
    \hfill & 0.1        & 3.00 (1.00) & 3.00 (1.00) & 3.00 (1.00) \\
    SNR=2 \ & 0.5 & 3.00 (1.00) & 3.00 (1.00) & 3.02 (0.99) \\
    \hfill & 0.9        & 2.94 (0.86) & 2.94 (0.86) & 3.06 (0.93) \\\hline
    \hfill & 0.1        & 3.00 (1.00) & 3.00 (1.00) & 3.00 (1.00) \\
    SNR=4 \ & 0.5 & 3.00 (1.00) & 3.00 (1.00) & 3.00 (1.00) \\
    \hfill & 0.9        & 3.06 (0.94) & 3.06 (0.94) & 3.01 (0.99) \\
    \hline
  \end{tabular}
\end{table}

\begin{table}[htbp]
  \centering
  \caption{Simulation results for Setting 1 with true rank $r_y=3$. The mean and the percentage of matching with the true rank (in parenthesis) of $r_y$ estimation over 300 simulation runs are presented. }\label{tab:ry_cv}
  \begin{tabular}{lrrrr}
    \hline
    & \multicolumn{1}{c}{$\rho$} & \multicolumn{1}{c}{NRRR} & \multicolumn{1}{c}{NRRS}\\\hline
    \hfill & 0.1        & 3.17 (0.92) & 3.04 (0.95) \\
    SNR=1 \ & 0.5 & 3.15 (0.92) & 3.05 (0.95) \\
    \hfill & 0.9        & 3.12 (0.93) & 3.05 (0.95) \\\hline
    \hfill & 0.1        & 3.01 (0.99) & 3.03 (0.98) \\
    SNR=2 \ & 0.5 & 3.00 (1.00) & 3.01 (0.99) \\
    \hfill & 0.9        & 3.05 (0.95) & 3.04 (0.96) \\\hline
    \hfill & 0.1        & 3.00 (1.00) & 3.01 (0.99) \\
    SNR=4 \ & 0.5 & 3.00 (1.00) & 3.01 (0.99) \\
    \hfill & 0.9        & 3.01 (0.99) & 3.01 (0.99) \\
    \hline
  \end{tabular}
\end{table}

%\begin{figure}
%\includegraphics[width=.32\textwidth]{figures/cv_boxplot_nestRRR_model1_s2n1_rX}\hfill
%\includegraphics[width=.32\textwidth]{figures/cv_boxplot_nestRRR_model1_s2n2_rX}\hfill
%\includegraphics[width=.32\textwidth]{figures/cv_boxplot_nestRRR_model1_s2n4_rX}\hfill
%\caption{Prediction error from model {\bf setting 1} by {\bf Cross-Validation}. From left to right are the plots with SNR being 1, 2, and 4. In each plot, the black boxplot is the one for $\rho=0.1$, the gray one is for $\rho=0.5$, and the white one is for $\rho=0.9$.}\label{fig:boxplot_cv}
%\end{figure}

\begin{figure}[htp]
\centering 
\subfigure[SNR = 1]{\includegraphics[width=0.32\textwidth]{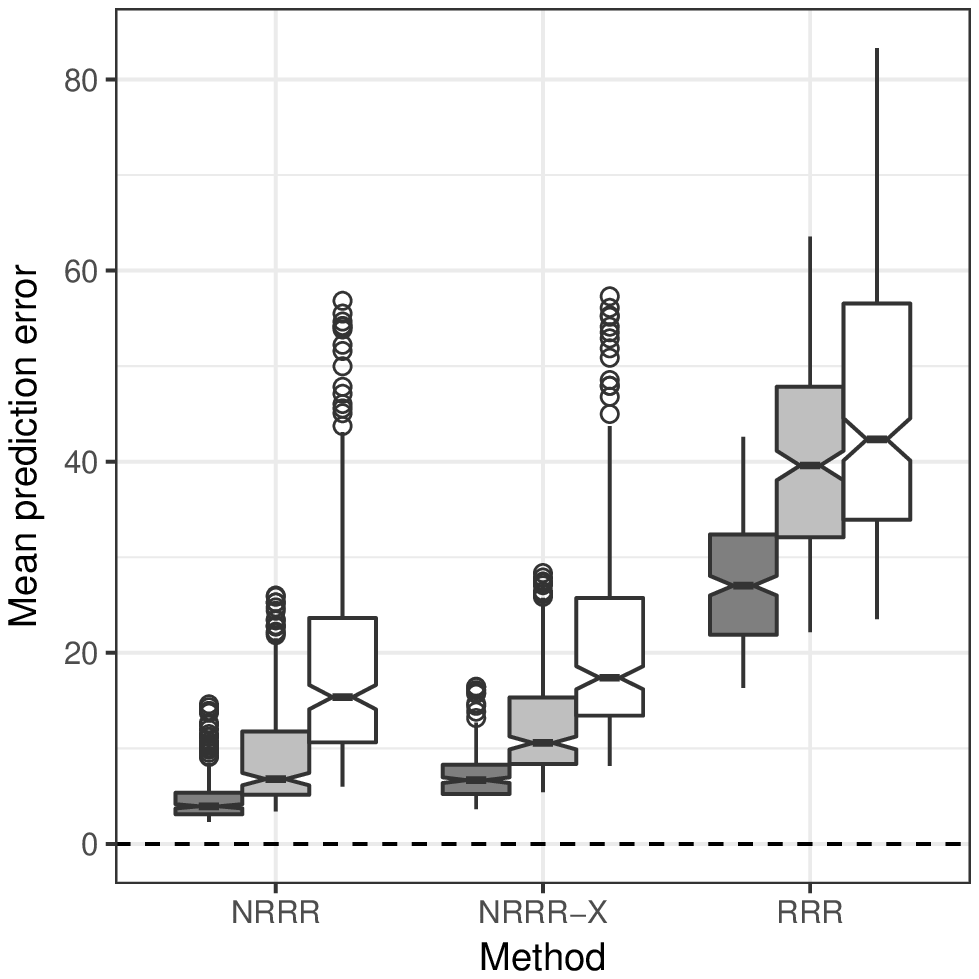}}
\subfigure[SNR = 2]{\includegraphics[width=0.32\textwidth]{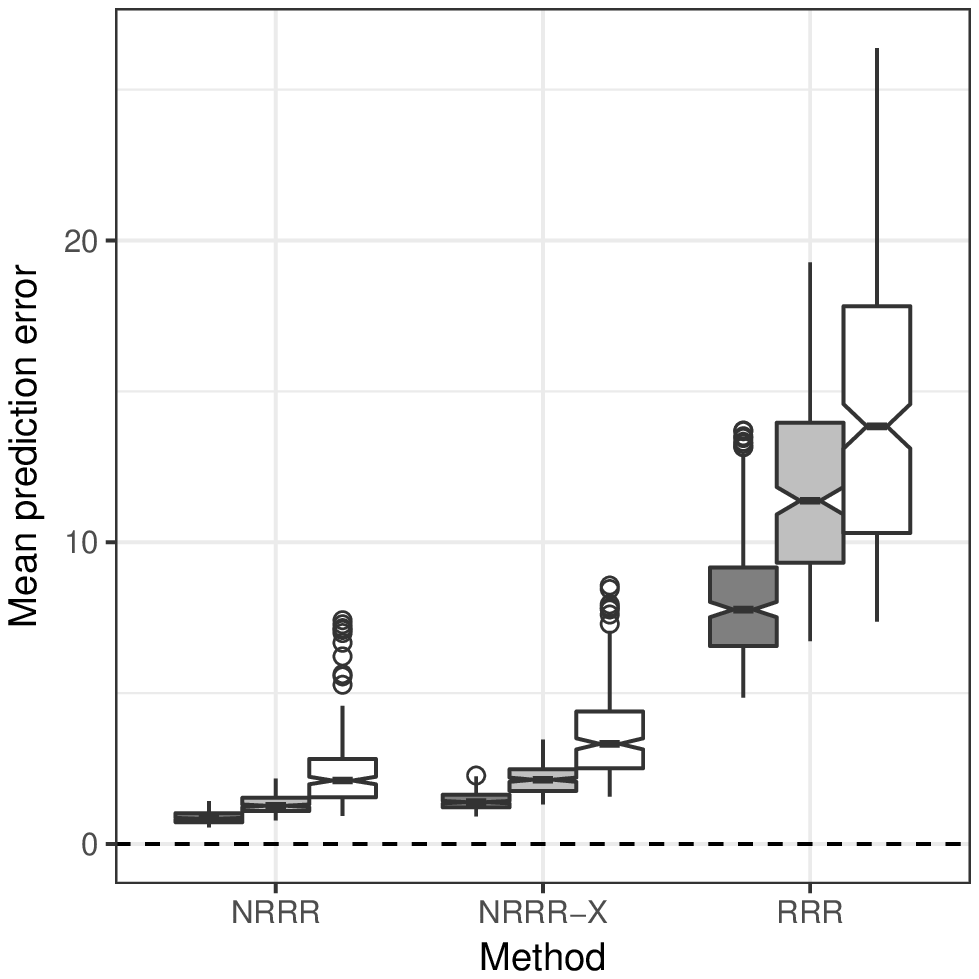}}
\subfigure[SNR = 4]{\includegraphics[width=0.32\textwidth]{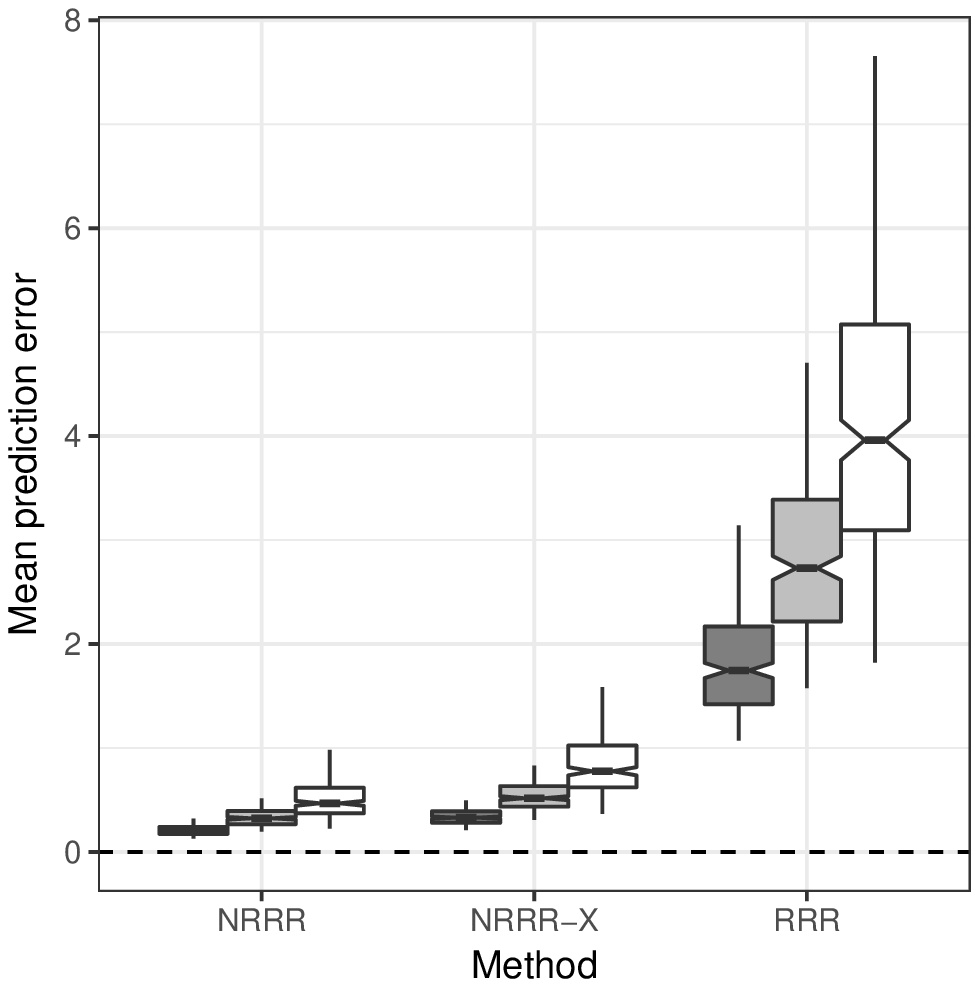}}
\caption{Boxplots of MSFPE from 300 simulation runs. From left to right are three plots with SNR being 1, 2, and 4. In each panel, each set of three boxplots for $\rho = 0.1, 0.5, 0.9$ is showing in black, grey and white colors from left to right.}\label{fig:boxplot2}
\end{figure}

%\clearpage
%\section*{D: Additional Graphs in Application Section}
%Following are the figures that display the temperature and electricity demand profiles of all the Mondays from 7/6/1997 to 3/31/2007 of Adelaide.

%\begin{figure}[htp]
%\subfigure[]{\includegraphics[width=.32\textwidth]{figures/mondaydemand}}
%\subfigure[]{\includegraphics[width=.32\textwidth]{figures/Mondaytempairport}}
%\subfigure[]{\includegraphics[width=.32\textwidth]{figures/Mondaytempkent}}
%\caption{Electricity demand in Adelaide (a), and temperature in Kent town (b) and airport (c). Plotted are the half-hourly observed profiles for all Mondays.}\label{Ade:Monday}
%\end{figure}

\newpage
\bibliographystyle{jasa3}
%\bibliography{nRRR}

\end{document}